\documentclass[a4paper,11pt]{article}
\usepackage{jcappub} % for details on the use of the package, please see the JINST-author-manual
\usepackage[dvipsnames]{xcolor}
\usepackage[normalem]{ulem}
\usepackage{subcaption}
\usepackage{hyperref}
\usepackage{ulem}
\usepackage{makecell}

\title{
Cosmic-ray anisotropy: sensitivity of methods and implications for KASCADE data
}

\author[a]{M.\,Yu.~Kuznetsov}
\author[a]{N.\,A.~Petrov}
\author[a, b*]{E.\,M.~Shinkevich}

\affiliation[a]{Institute for Nuclear Research of the Russian Academy of Sciences, \\
60th October Anniversary Prospect 7a, Moscow 117312, Russia}
\affiliation[b]{Lomonosov Moscow State University, Leninskie Gory 1, Moscow, 119991, Russia}
\emailAdd{shinkevich.em20@physics.msu.ru}

\abstract{
We study the problem of measuring the anisotropy of high-energy cosmic rays across all angular scales. The limited field of view and non-uniform exposure of ground-based cosmic-ray experiments reduce their sensitivity to real anisotropy of cosmic-ray arrival directions. A widely used signature of anisotropy --- a dipole of the flux expansion over the right ascension --- provides very limited understanding of the underlying physics of cosmic-ray origin and propagation. In this study we test two other methods: the angular power spectrum and the autocorrelation function for sensitivity to possible medium- and small-scale anisotropies of cosmic-ray flux. We find that the autocorrelation function is the most sensitive estimator for an underlying physical anisotropy in most of the models tested, while the angular power spectrum can provide additional knowledge about cosmic-ray flux properties, when the anisotropy is strong enough. As a test of our findings, we apply these methods to $10\%$ sample of the KASCADE experiment public data. Namely, we consider all-particle set of events and sets of individual mass groups classified by a convolutional neural network. We find an indication of anisotropy at $> 2.5 \sigma$ level at $\sim 10^\circ$ angular scale for the iron nuclei mass group at $E \gtrsim 20$~PeV with both the angular power spectrum and the autocorrelation methods.
}


\begin{document}
\maketitle
\flushbottom

\section{Introduction}
\label{sec:intro}

Even a century after their discovery, our understanding of the nature of cosmic rays (CRs) remains incomplete~\cite{Gabici_2019, Kachelriess:2019oqu}. In particular, the origin of particles with energies greater than 1~PeV has not yet been definitively established. It is generally assumed that up to energies of approximately 3--4~PeV for protons and 80~PeV for iron nuclei, CRs are of galactic origin and are effectively confined by the magnetic field of our Galaxy. For a long time, galactic supernovae were considered the primary candidates to accelerate particles to such extreme  energies~\cite{1961PThPS..20....1G}. However, recent observations of PeVatron gamma-ray sources by the LHAASO observatory have revealed only a partial correlation with supernovae, calling this model into question~\cite{LHAASO:2021gok}.

In this regard, the arrival direction anisotropy of high-energy CRs can be a useful tool to understand the nature of their sources and the mechanisms of particle diffusion within the interstellar medium~\cite{Ahlers:2016rox}. The presence of a large-scale dipole anisotropy, predicted within the framework of the diffusive particle propagation model~\cite{Gabici_2019}, has been confirmed by several experiments. In particular, at energies greater than $100$~TeV, significant anisotropy was found by EAS-TOP~\cite{EAS-TOP:2009nld}, Tibet AS$\gamma$~\cite{Amenomori:2017jbv}, ARGO-YBJ~\cite{ARGO-YBJ:2018zoa}, KASCADE-Grande~\cite{Apel:2019afz}, IceTop~\cite{IceCube:2012vve}, IceCube~\cite{IceCube:2016biq, IceCube:2024pnx}, and LHAASO~\cite{Gao:2023jlz}. However, the amplitude of such a dipole was significantly smaller than the predictions of the standard diffusive model~\cite{Hillas:2005cs, Gabici_2019}, and in addition, unexpected medium-scale anisotropies with angular sizes of $\gtrsim 10^\circ$ have also been observed~\cite{IceCube:2016biq, IceCube:2024pnx}. Another unexpected feature of the medium-scale was found in the data of the KASCADE-Grande experiment~\cite{Ahlers:2019gdc}. Although the latter observations were explained by the turbulent magnetic field that redistributes the initial dipole anisotropy to smaller angular scales~\cite{PhysRevLett.112.021101, Ahlers:2016rox}, a more detailed study of anisotropies at different angular scales is plausible for a better understanding of the origin and propagation of CRs. In particular, one could expect the anisotropy to be enhanced at distinct CR magnetic rigidities~\cite{PhysRevLett.112.021101}. In turn, this effect could be searched for by studying the anisotropy of separate CR mass groups. Several studies of this type have recently been performed~\cite{Kostunin:2021eyp, He:2025N5, Lebedev_2026}, though medium- and small-scale anisotropies have not been considered systematically.

As the measurements of medium- and small-scale anisotropy of high-energy CRs would yield additional knowledge on their origin and propagation, it is interesting to study the relative sensitivity of different methods of anisotropy detection. In the present study, we perform a comparative analysis of two key methods for searching for such anisotropy: the autocorrelation function and the angular power spectrum. In particular, we examine their sensitivity to different anisotropy models under the condition of a limited detector field of view. As an illustration, we apply the investigated methods of anisotropy study to the data of the KASCADE experiment~\cite{KASCADE:2003swk}, publicly provided by the KCDC service~\cite{Haungs_2018}. Namely, we use a $10\%$ sample of the data classified with the convolutional neural network into five primary particle mass groups (p, He, C, Si, and Fe)~\cite{Kuznetsov:2023kss, Kuznetsov:2023pvo}.

The paper is organized as follows: Sec.~\ref{sec:method} describes the methods used to analyze anisotropy on large, small, and intermediate angular scales. Sec.~\ref{sec:sens} presents a comparison of the sensitivity of the autocorrelation function and the angular power spectrum for three different anisotropy models. The KASCADE experiment and the dataset used in this study are described in Sec.~\ref{sec:expMC}. The application of the anisotropy search methods to the KASCADE data is presented in Sec.~\ref{sec:results} for all-particle event set as well as for sets of different primary mass groups. Finally, Sec.~\ref{sec:conclusion} summarizes the main results of this work.
%%%%%%%%%%%%%%%%%%%%%%%%%%%%%%%%%%%%%%%%%%%%%%%%%%%%%%

\section{Methods of anisotropy search}
\label{sec:method}

\subsection{Large-scale anisotropy}
We begin with the most traditional analysis: a one-dimensional search for a large-scale anisotropy in the distribution of CR events. The standard method is the Rayleigh analysis of the first Fourier harmonic in the right ascension. However, this method is highly susceptible to systematic errors~\cite{Linsley:1975kp}. In this work, the East-West method~\cite{Bonino:2011nx} is applied to analyze large-scale anisotropy. The key advantage of this approach lies in its invariance with respect to detector exposure. This eliminates the influence of diurnal intensity variations and instrumental effects on the estimation of the dipole component.

In this method, for $N$ events with local sidereal arrival times $t_i$, the dipole parameters are estimated by means of a harmonic analysis of the flux difference between the eastern and western hemispheres. The coefficients of the first-harmonic expansion are defined as
\begin{equation}
\begin{split}
a &= \frac{2}{N} \sum_{i=1}^N \cos(t_i + \zeta_i), \\
b &= \frac{2}{N} \sum_{i=1}^N \sin(t_i + \zeta_i),
\end{split}
\label{eq:eastwest}
\end{equation}
where the phase shift $\zeta_i$ takes the values $0$ and $\pi$ for events detected in the local eastern and western hemispheres, respectively. These hemispheres are defined relative to the local meridian in the horizontal coordinate system (azimuth $0^\circ < \phi < 180^\circ$ for East and $180^\circ < \phi < 360^\circ$ for West). From these coefficients, the amplitude of the dipole projection onto the equatorial plane, $\hat{D}_{\perp}$, and its phase, namely the right ascension $\hat{\alpha}_d$, are reconstructed as
\begin{equation}
\begin{split}
\hat{D}_{\perp} &= \frac{\pi}{2\langle\sin\theta\rangle}\sqrt{a^2+b^2}, \\
\hat{\alpha}_d &= \frac{\pi}{2} + \arctan\left(\frac{b}{a}\right),
\end{split}
\end{equation}
where $\langle\sin\theta\rangle$ denotes the mean value of the sine of the zenith angle of the detected events.

An important feature of the East-West method is that the statistical significance of the obtained result can be calculated analytically.
Under the null hypothesis of an isotropic distribution of particle arrival directions, the dipole amplitude follows a Rayleigh distribution with parameter
\begin{equation}
\sigma = \frac{\pi}{2 \langle \sin\theta \rangle} \sqrt{\frac{2}{N}},
\end{equation}
while the phase is uniformly distributed. The probability that the observed amplitude $\hat{D}_{\perp}$ is a fluctuation of the isotropic background is given by
\begin{equation}
p(D_{\perp} > \hat{D}_{\perp}) = \exp\left(-\frac{\hat{D}_{\perp}^2}{2\sigma^2}\right).
\end{equation}

\subsection{Small-scale and medium-scale anisotropy}

Currently, a wide range of methods is available for evaluating small- and medium-scale anisotropy in the CR distribution on the celestial sphere. Among the most widely used methods are the analysis of the two-point correlation function~\footnote{In what follows, we call it the autocorrelation function, according to the tradition of CR anisotropy studies.} and the multipole expansion of the angular distribution over spherical harmonics. %which allows for determining the contribution of individual multipole components to the observed anisotropy.

The spatial autocorrelation function on the sphere measures the degree of event clustering or the excess probability of finding two objects at a given distance from each other. The standard estimator for this function is based on the counting of the number of event pairs separated by the given angle $\psi$~\cite{1980lssu.book.....P}:
\begin{equation}
w(\psi) = \frac{DD(\psi) }{RR(\psi)} - 1,
\label{eq:autocorr_std}
\end{equation}
where $DD(\psi)$ is the number of event pairs in the analyzed dataset and $RR(\psi)$ is the number of event pairs in the Monte Carlo sample corresponding to an isotropic distribution.
This is the normalized version of a simple counting of the number of pairs that is traditionally used in CR studies~\cite{PierreAuger:2022axr}.
%It should be noted that since the primary focus of this study is the direct testing of the cosmic-ray isotropy hypothesis,
In this study we employ the autocorrelation function that is cumulative in angle $\psi$,
%Due to the integration effect, where
i.e. all pairs in Eq.~(\ref{eq:autocorr_std}) are counted within an angular scale $\psi$. This approach trades off the specificity of the differential autocorrelation function to correlation on a peculiar angular scale for an improvement of the sensitivity to correlation on any scale.

\begin{figure}[!ht]
    \centering
    \begin{subfigure}[b]{0.48\textwidth}
        \centering
        \includegraphics[width=\textwidth]{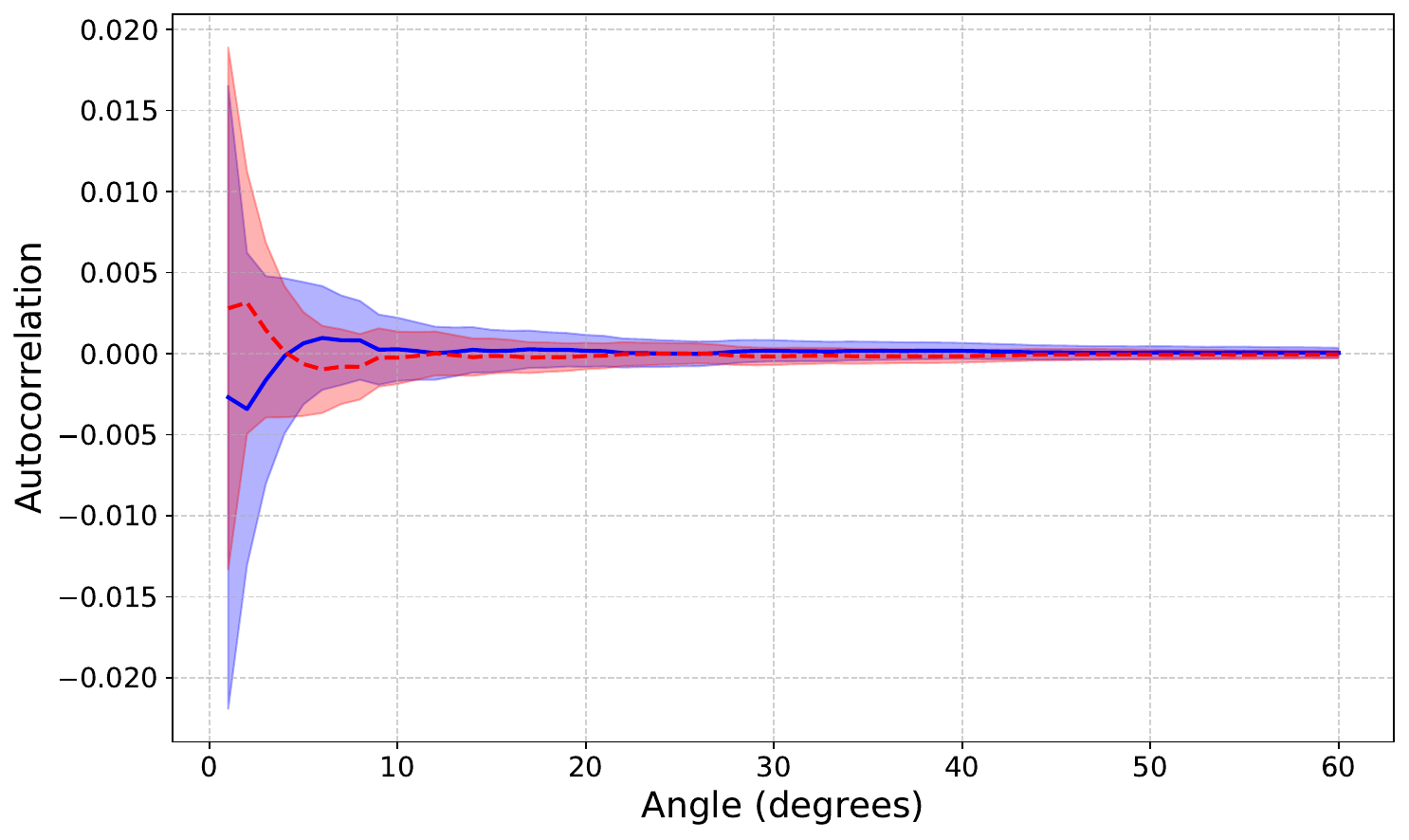}
        \caption{Uniform exposure}
        \label{fig:AC_LS_uni}
    \end{subfigure}
    \hfill
    \begin{subfigure}[b]{0.48\textwidth}
        \centering
        \includegraphics[width=\textwidth]{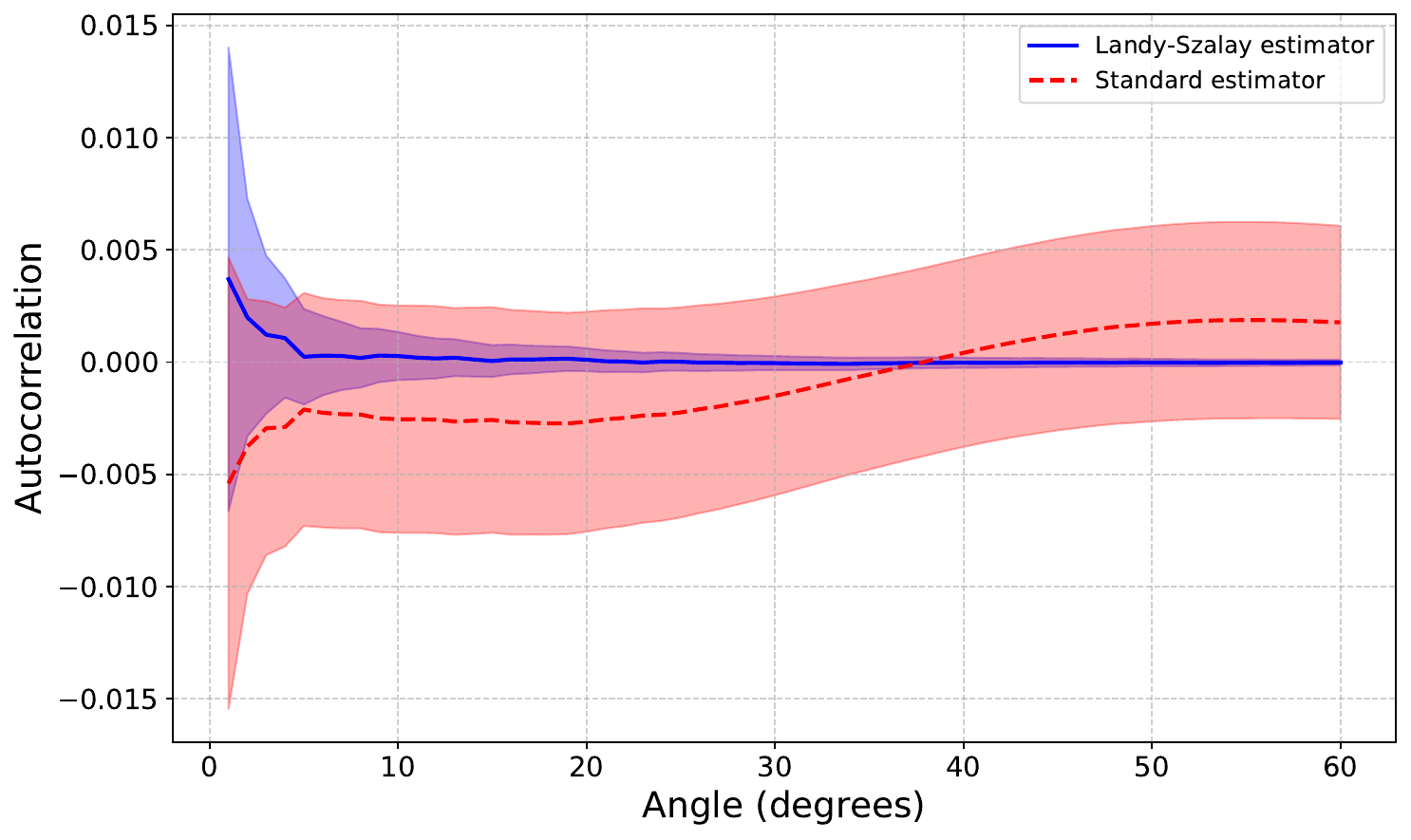}
        \caption{KASCADE exposure}
        \label{fig:AC_LS_exp}
    \end{subfigure}
    
    \caption{Comparison of the mean values of the standard and Landy-Szalay estimators for the cumulative autocorrelation function, based on 10,000 isotropic datasets with 10,000 events each. The shaded bands represent the $\pm \sigma$ intervals. Panel (a) shows the results for a uniform full-sky exposure, while panel (b) corresponds to datasets generated from the KASCADE geometric exposure map (see Sec.~\ref{sec:sens} for details).}
    \label{AC_LS-vs-Std}
\end{figure}

The improved estimator for autocorrelation function was proposed  by Landy and Szalay~\cite{1993ApJ...412...64L}:
\begin{equation}
w(\psi) = \frac{DD(\psi) - 2 DR(\psi) + RR(\psi)}{RR(\psi)}, \label{eq:autocorr}
\end{equation}
where $DR(\psi)$ is the number of cross-pairs between data and Monte Carlo sets. This estimator handles exposure edge effects significantly better than the standard estimator~(\ref{eq:autocorr_std}), which is crucial given the limited exposure of ground-based CR detectors. Although both autocorrelation estimators are unbiased, the variance of the Landy-Szalay method is significantly smaller than that of the standard estimator. This reduced variance is particularly important for evaluating the significance of deviations from isotropy. Additionally, the Landy-Szalay estimator converges more rapidly to the true value. These differences are demonstrated in Fig.~\ref{AC_LS-vs-Std}, which is based on 10,000 isotropic sets generated for a uniform exposure map and the KASCADE geometric exposure map (see Sec.~\ref{sec:sens} for details). The autocorrelation function was calculated using both the standard and Landy-Szalay estimators for each dataset of 10,000 events. One can see that the Landy-Szalay estimator is much more precise in the case of a limited field of view for the experiment.

The multipole analysis is based on the expansion of the relative CR flux on the sphere in terms of spherical harmonics with coefficients $a_{lm}$:
\begin{equation}
\delta I(\vec{n}) = \sum_{l=1}^{\infty}\sum_{m=-l}^{l}a_{lm}Y_{lm}(\vec{n}).
\end{equation}
%In this study, to test the hypothesis of an isotropic distribution, we investigate the angular power spectrum:
Given the set of $a_{lm}$, it is convenient to estimate the average level of the anisotropy at a given angular scale using angular power spectrum:
\begin{equation}
C_l = \frac{1}{2l+1} \sum_{m=-l}^{l}|a_{lm}|^2,
\label{eq:std_Cl}
\end{equation}
%where anisotropy manifests as elevated values of 
where the given $C_l$ characterizes anisotropy at the angular scale $\sim \pi/l$.
%However, it should be noted that the KASCADE experiment lacks 
For the experiments with non-full and non-uniform sky coverage the power spectrum derived directly from the observed flux map
is a convolution of the power spectrum of the exposure
and the power spectrum of the true flux.
In this case, it is referred to as the pseudo-power spectrum $\tilde{C}_l$ that suffers from mixing between adjacent multipoles $l$.
However, these effects can be corrected for. In this study, we use the PolSpice software package~\cite{2004MNRAS.350..914C, 2001ApJ...548L.115S} for this purpose. This code utilizes the relationship between the power spectrum and the spatial correlation function via Legendre polynomials:
\begin{equation}
\tilde{w}(\psi)= \frac{1}{4\pi}\sum_{l=0}^{\infty}(2l+1)\tilde{C}_l P_l(\cos{\psi}).
\end{equation}
The correlation function $\tilde{w}(\psi)$, evaluated from the pseudo-power spectrum, is corrected in the spatial domain and then inverted back into the true angular power spectrum $\bar{C}_l$:
\begin{equation}
\bar{C}_l = 2\pi\int_{-1}^{1}\bar{w}(\psi) P_l(\cos{\psi}) d(\cos{\psi}),
\end{equation}
where $\bar{w}(\psi)$ is the corrected correlation function. This allows us to partially eliminate the correlations between separate modes. While low-order multipoles ($l=1, 2$) cannot be unambiguously reconstructed in this analysis due to non-full sky coverage, their relatively large values can have impact on higher multipoles (particularly $l=3, 4$), potentially introducing distortions.

%\section{Sensitivity of the autocorrelation function and the power spectrum}
%\section{Sensitivity of various anisotropy signatures}
\section{Test of sensitivity of methods}
\label{sec:sens}
In this section we test the relative sensitivity of small- and medium-scale anisotropy estimators to the underlying anisotropy of the CR flux using several flux models. We consider the isotropic distribution of the flux as a null hypothesis. Namely, we use isotropic Monte Carlo datasets taking into account the exposure of the detector. In this part of the study we consider the idealized model of the exposure that is uniform over the azimuthal angle and have the following distribution over the zenith angle:
\begin{equation}\begin{split}
\frac{dN}{d\cos\theta} \propto \cos{\theta},
%\quad \phi \sim \mathcal{U}(0, 2\pi),
\end{split}\end{equation}
In terms of equatorial coordinates, this translates to the so called geometric relative exposure~\cite{Sommers:2000us}, which is determined by the detector latitude $a_0$ and the maximum zenith angle $\theta_{max}$:
\begin{equation}
\omega(\delta) \propto \cos(a_0)\cos(\delta)\sin(\alpha_m) + \alpha_m\sin(a_0)\sin(\delta),
\end{equation}
where $\alpha_{m}$ is given by:
\begin{displaymath}
\alpha_{m} = 
    \begin{cases}
        0, &\text{if} \ \xi > 1\\
        \pi, &\text{if}\ \xi < -1\\
        \arccos(\xi) &\text{otherwise}
    \end{cases}
\end{displaymath}

For the construction and analysis of sky maps, we use the HEALPix software package, specifically its Python implementation, the healpy library\cite{Gorski_2005, Zonca2019}. The $N_{\text{side}}$ parameter was set to 128, corresponding to 196,608 pixels over the entire sky. This value of $N_{\text{side}}$ corresponds to an angular pixel size of approximately $0.5^\circ$, which is significantly smaller than the angular scale of the expected CR anisotropy at these energies. It is worth noting that this choice of $N_{\text{side}}$ also enables a more accurate  reconstruction of the exposure map using the time-scrambling method, which is discussed in detail in Sec.~\ref{sec:expMC}.

To estimate the statistical significance of the detected anisotropy, the local significance, or pre-trial $p$-value, is calculated. It is determined independently for each angular scale $\psi$ (in the autocorrelation analysis) or each multipole $l$ (in the harmonic analysis) as the fraction of simulated isotropic datasets in which the anisotropy amplitude exceeds the experimentally observed value. We want to stress that in this study we do not account for deficits with respect to isotropy for the $p$-value. However, since the search is conducted over a broad parameter space, the look-elsewhere effect must be considered. For a robust interpretation of the results, the global significance, or post-trial $p$-value, is introduced, which accounts for the probability of a large enough statistical fluctuation occurring anywhere within the entire interval of parameters. Namely, this quantity is evaluated
%using Monte Carlo realizations
as the fraction of simulated isotropic datasets where, at any point within the considered range, the local $p$-value is equal or lower than the minimum pre-trial $p$-value found in the experiment. The range of parameters to scan is determined by the characteristic size of the detector's field of view (FoV). In particular, for the KASCADE experiment the characteristic scale corresponds to approximately $60^\circ$ if one uses a zenith angle cut $\theta < 30^\circ$. Therefore, the post-trial $p$-value is computed over the angular range $\psi < 60^\circ$ and the multipole range $l > 3$.

To ensure an accurate comparison between autocorrelation and angular power spectrum, we additionally introduce a cumulative statistic for the angular power spectrum. The $C_l$ values were calculated up to $l_{max} = 180$ with a bin width of $\Delta l = 4$. We define the cumulative power spectrum by analogy with the autocorrelation function, where the summation is performed in an angular range limited from above, starting from the smallest angular scales (highest multipoles):
\begin{equation}
C_i^{cum} = \sum_{j=i}^{N_{bin}} C_j^{bin},
\label{eq:cl_sum}
\end{equation}
where $i$ is the bin index, and $N_{bin}$ is the total number of bins.

We evaluate the sensitivity of both autocorrelation and multipole methods to different anisotropy scenarios for the observation in a CR experiment located at the middle latitudes (for certainty, we use the coordinates of the KASCADE experiment) and assume the geometric exposure of the detector with $\theta < 30^\circ$. We generate 10,000 isotropic sets (following the exposure map) for background estimation, which allows us to resolve post-trial $p$-values down to the $10^{-3} - 10^{-4}$ level. For comparison, we used the median pre-trial and post-trial $p$-values evaluated over $N_{map}=10$ map realizations for each flux model described further (except the separate source model, where the results do not depend on realization). We sample 1000 event sets per map; each set consists of 10,000 events. The anisotropic component of the flux map is modeled as:
\begin{equation}
\delta I(\vec{n}) = f(\vec{n}),
\end{equation}
yielding the total intensity:
\begin{equation}
I(\vec{n}) = 1 + \epsilon \frac{f(\vec{n})}{\max(f(\vec{n}))},
\label{eq:anisotropy}
\end{equation}
where $\epsilon$ defines the level of anisotropy. The parameter $\epsilon$ is chosen to yield a detectable anisotropy  (on a post-trial $p$-value level of $10^{-2} - 10^{-3}$) given the limited number of events. We convolve the anisotropic flux map~\eqref{eq:anisotropy} with the geometric exposure of the experiment, normalize it appropriately and use the resulting map as probability density function to sample the anisotropic sets of events.

%\begin{enumerate}
%\item \textbf{Gaussian power spectrum model.}
\subsection{Gaussian power spectrum model}
\label{sec:gaussian_Cl}
\begin{figure}[!ht]
	\centering
    \includegraphics[width=0.9 \linewidth]{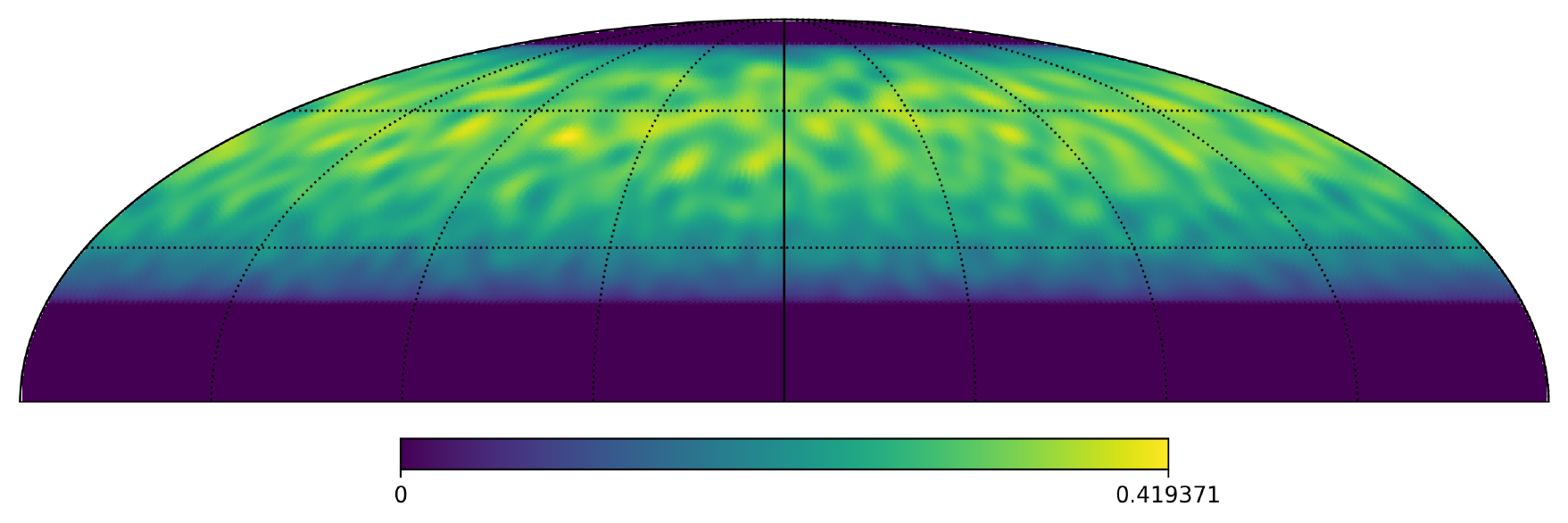}
    \caption{The example of the intensity map with a power spectrum following Eq.~\eqref{eq:gauss} ($\sigma = 10, l_0 = 30$) and taking into account the detector exposure.}
    \label{fig:gaussian_cl}
\end{figure}

First, we consider the toy model of the anisotropy, defined in terms of the angular power spectrum:
\begin{equation}
C_l = \exp\left(-\frac{(l - l_0)^2}{2\sigma^2}\right),
\label{eq:gauss}
\end{equation}
where $l_0$ and $\sigma$ are free parameters. The example of particular flux map realization for this model is shown in Fig.~\ref{fig:gaussian_cl}. By design, this model is more suitable for measurement with the power spectrum method, as it isolates a specific multipole. However, this model can also be utilized to investigate the extent to which power spectrum smearing affects the sensitivity of the anisotropic signatures. Therefore, we study the sensitivity of this model across various values of the parameters $\sigma$ and $l_0$.

\begin{figure}[!ht]
    \centering
    \begin{subfigure}[b]{0.48\textwidth}
        \centering
        \includegraphics[width=\textwidth]{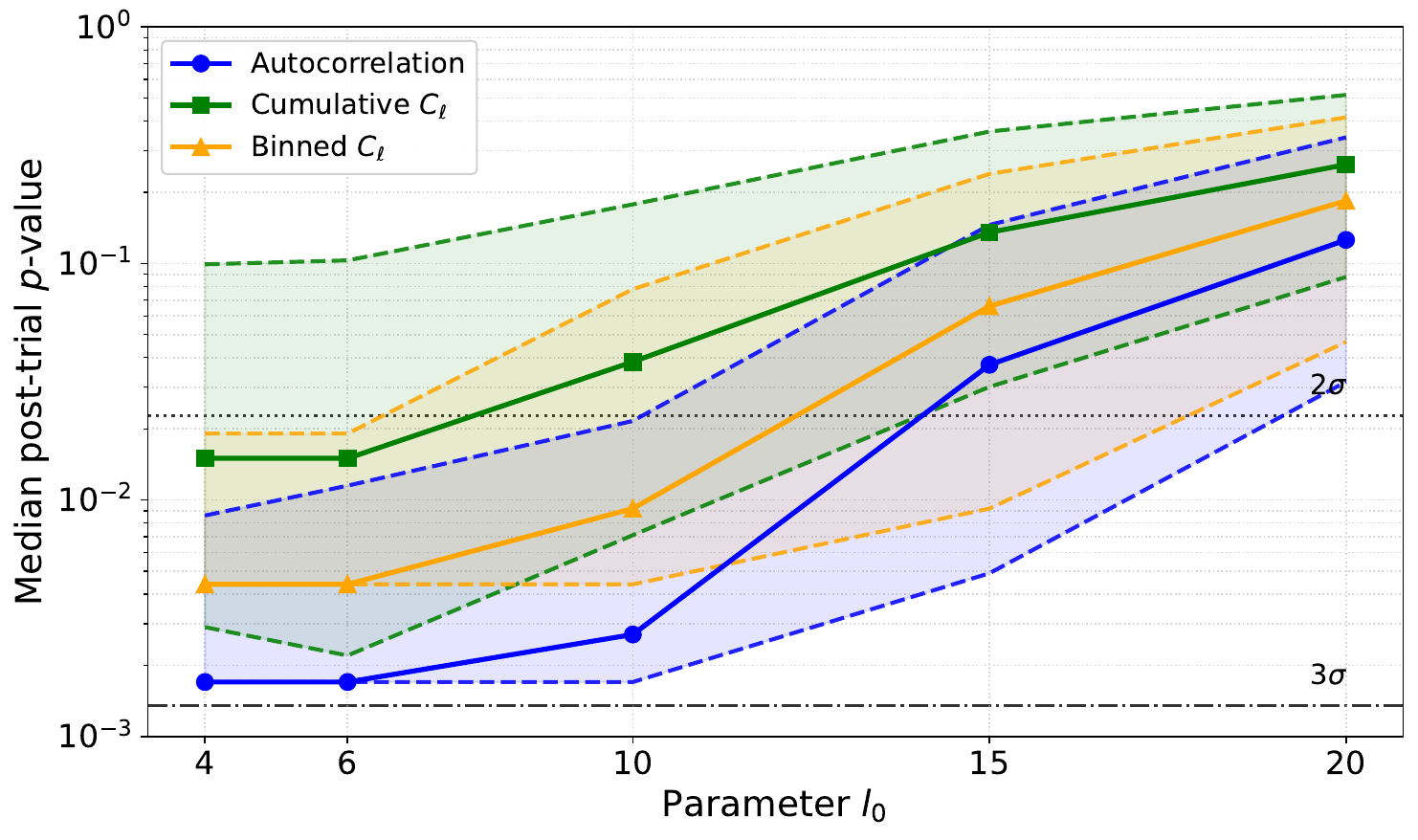}
        \caption{$\sigma=5.5$}
    \end{subfigure}
    \hfill
    \begin{subfigure}[b]{0.48\textwidth}
        \centering
        \includegraphics[width=\textwidth]{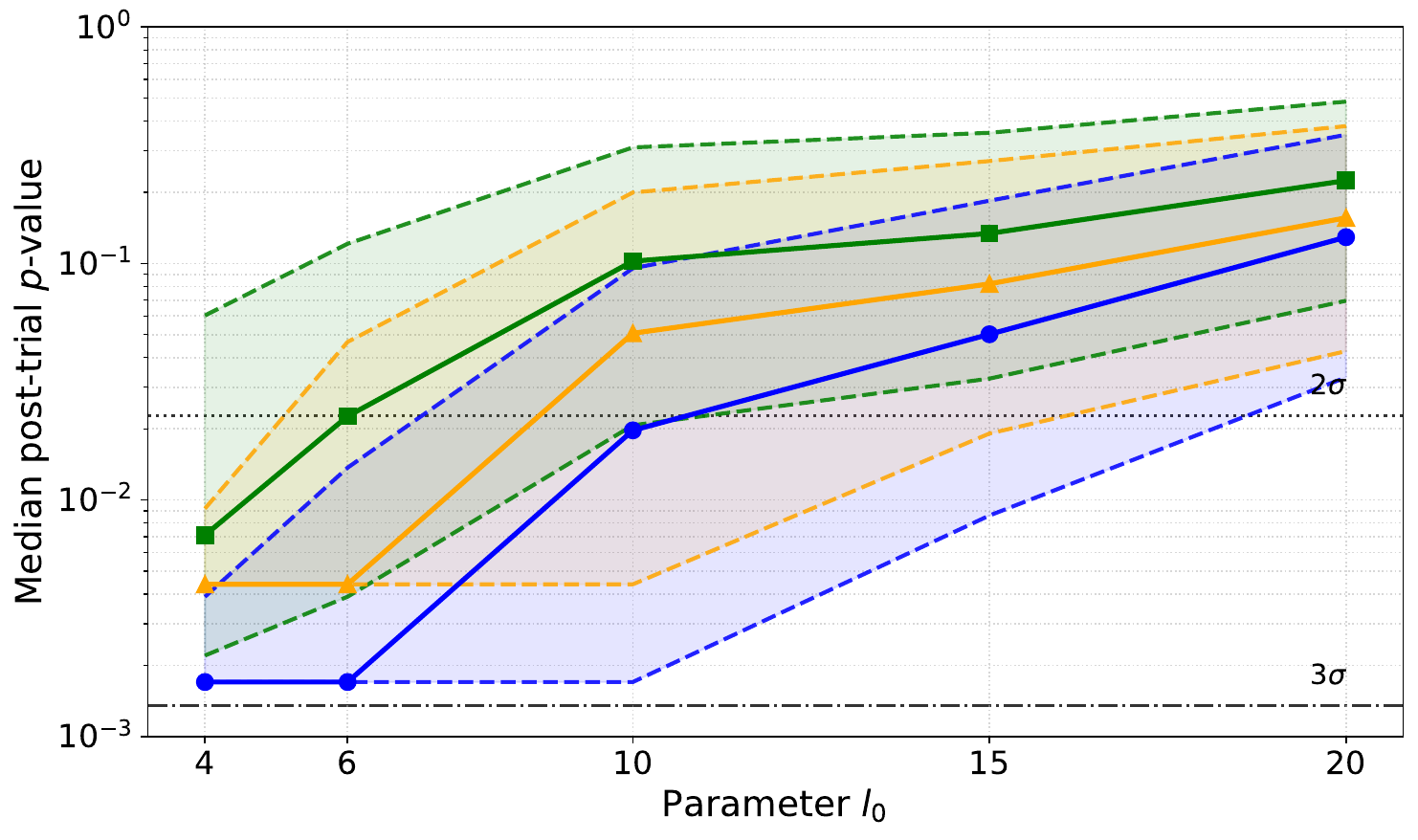}
        \caption{$\sigma=6$}
    \end{subfigure}
    
    \vspace{0.5cm}
    
    \begin{subfigure}[b]{0.48\textwidth}
        \centering
        \includegraphics[width=\textwidth]{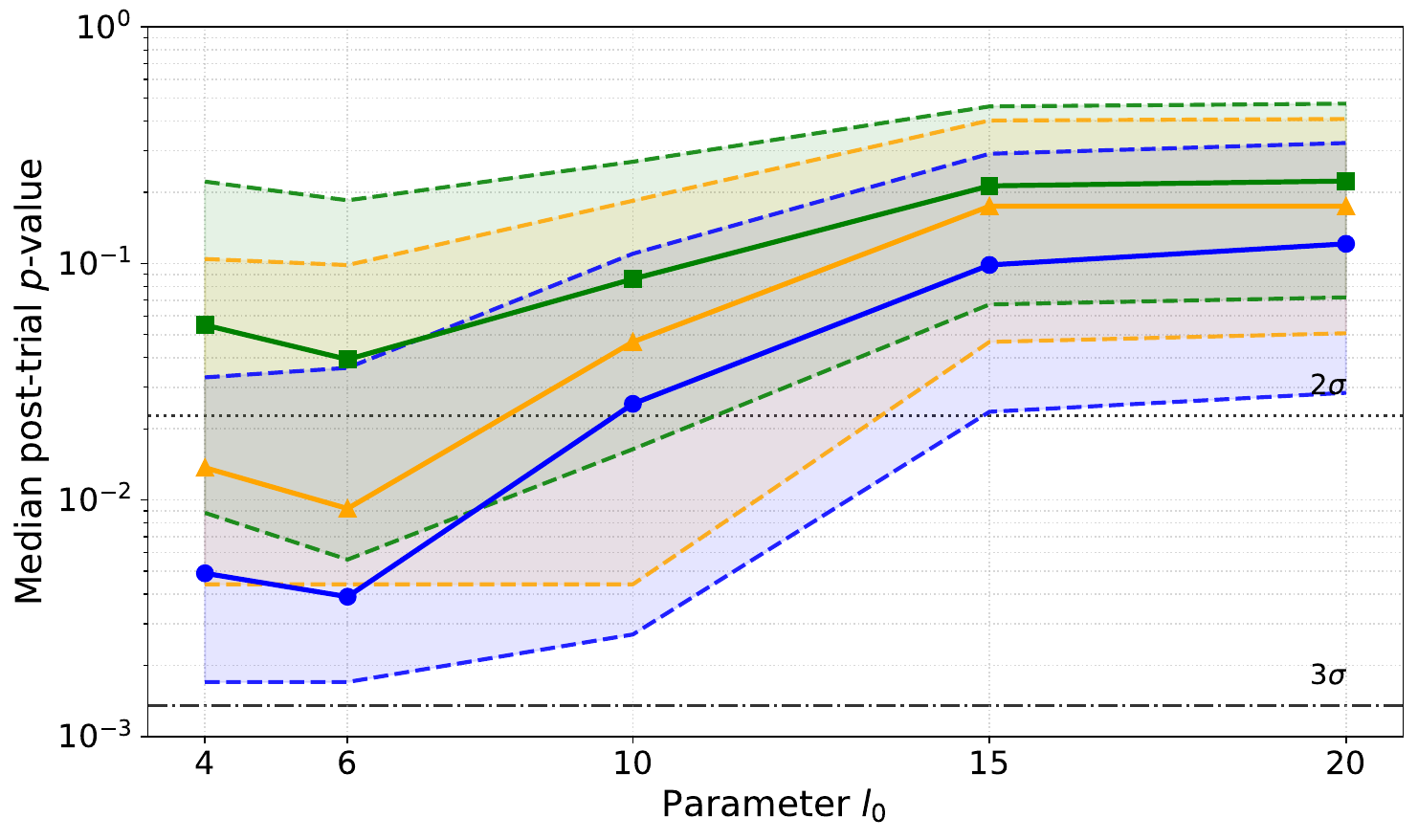}
        \caption{$\sigma=7$}
    \end{subfigure}
    \hfill
    \begin{subfigure}[b]{0.48\textwidth}
        \centering
        \includegraphics[width=\textwidth]{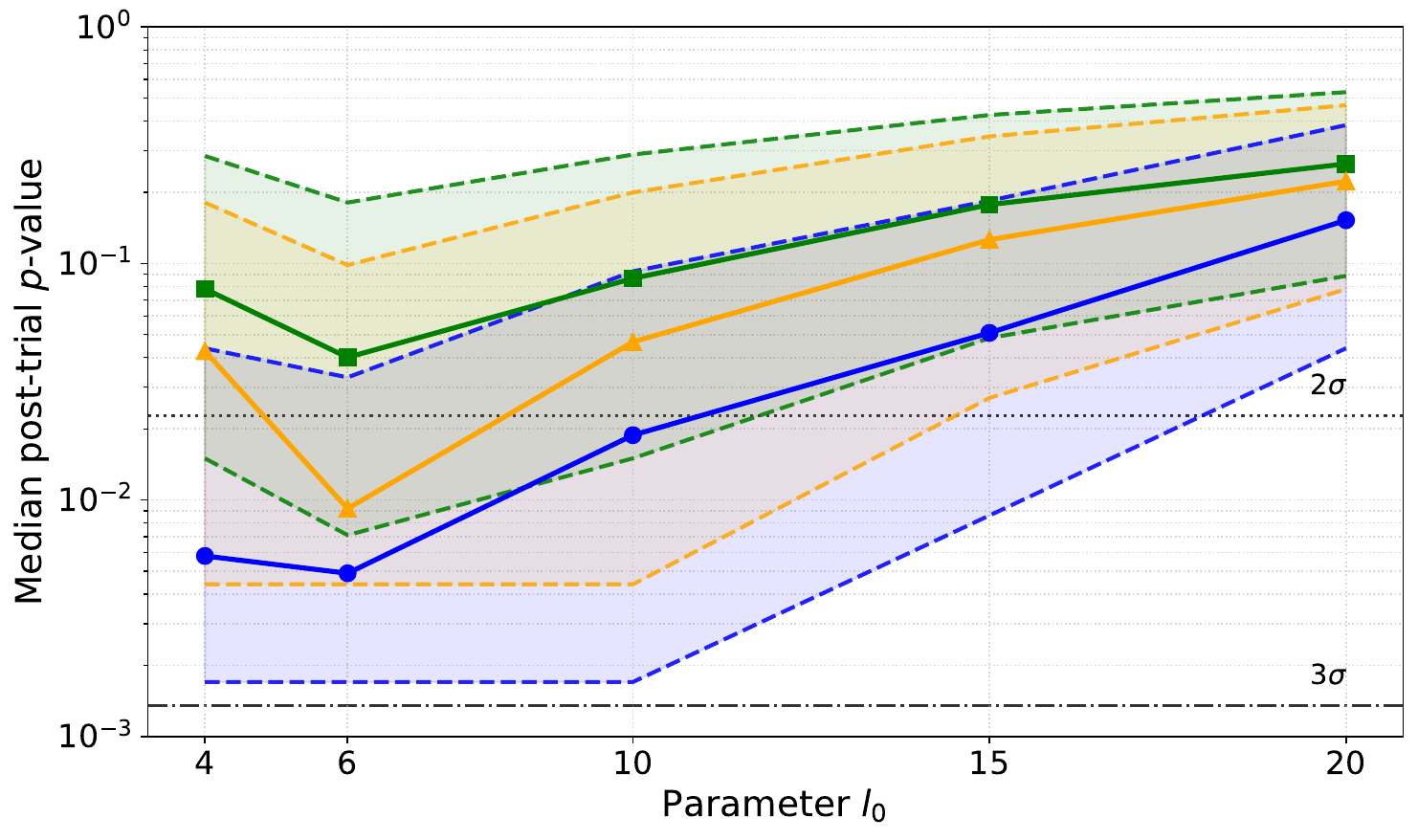}
        \caption{$\sigma=7.5$}
    \end{subfigure}
    
    \caption{Comparison of the median post-trial $p$-value between the autocorrelation function and the power spectrum for the model of Eq.~\eqref{eq:gauss} with varying $l_0$ at fixed $\sigma$. The shaded bands denote the range around the median: from the 25th to the 75th percentile.
}
    \label{fig:Ac-vs-Cl_gaussian_cl_p1}
\end{figure}

First, we consider the scenario with $\epsilon = 0.3$ in Eq.~\eqref{eq:anisotropy}, for which both methods are insensitive to anisotropy at high multipoles and, consequently, small angular scales. At the same time, for this anisotropy level, both methods demonstrate high sensitivity to medium scales and correspondingly low multipoles (see Fig.~\ref{fig:Ac-vs-Cl_gaussian_cl_p1}). However, a comparison of their sensitivities reveals that the autocorrelation method slightly outperforms the power spectrum in terms of the post-trial $p$-value.

\begin{figure}[!ht]
    \centering
    \begin{subfigure}[b]{0.48\textwidth}
        \centering
        \includegraphics[width=\textwidth]{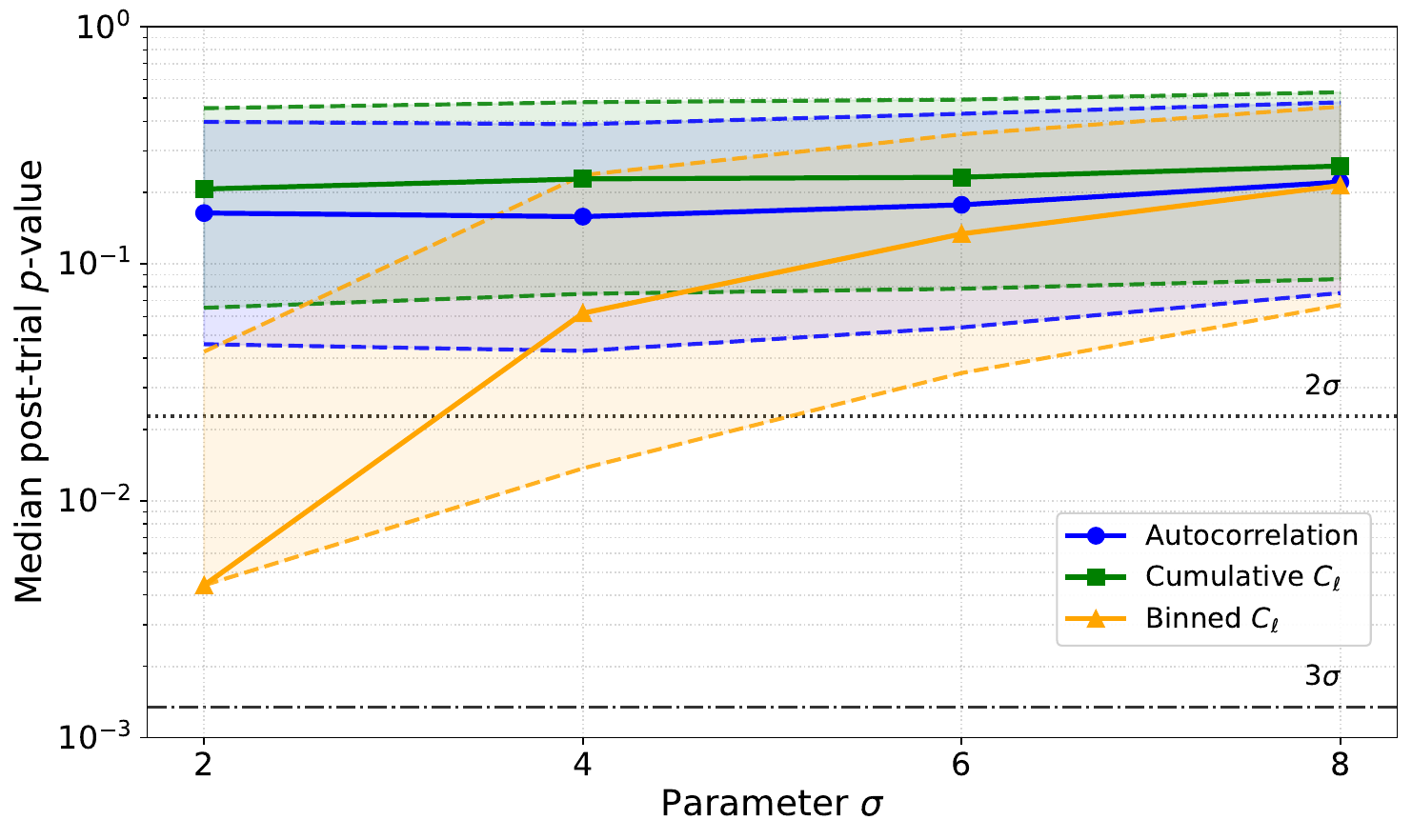}
        \caption{$l_0=30$}
    \end{subfigure}
    \hfill
    \begin{subfigure}[b]{0.48\textwidth}
        \centering
        \includegraphics[width=\textwidth]{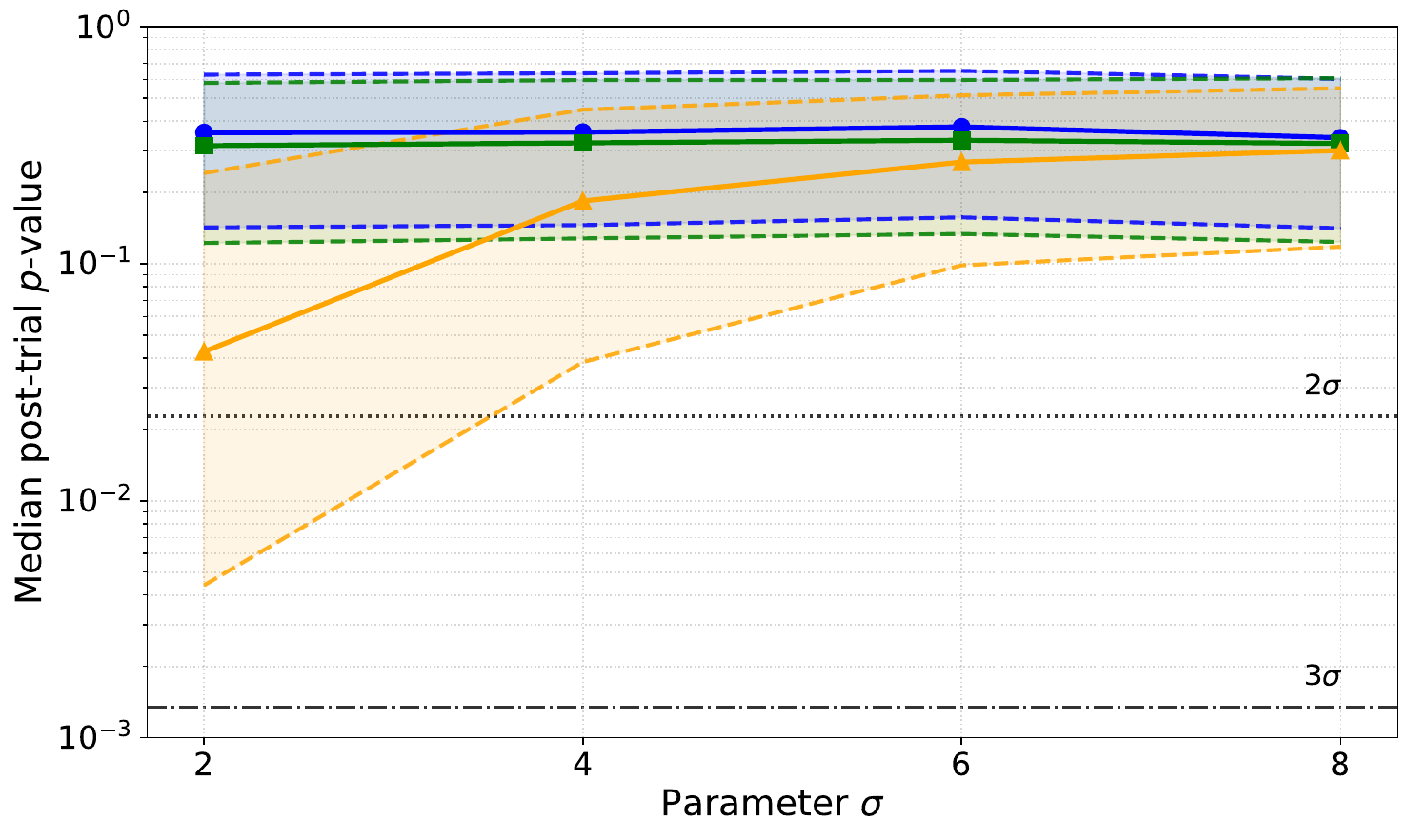}
        \caption{$l_0=50$}
    \end{subfigure}

    \caption{Comparison of the median post-trial $p$-value between the autocorrelation function and the power spectrum for the model in Eq.~\eqref{eq:gauss} with varying $\sigma$ at fixed $l_0$. The shaded bands denote the range around the median: from the 25th to the 75th percentile.}
    \label{fig:Ac-vs-Cl_gaussian_cl_p2}
\end{figure}

To study the sensitivity of the methods at smaller angular scales and higher multipoles we slightly increased the anisotropy amplitude, setting $\epsilon = 0.35$ in Eq.~\eqref{eq:anisotropy}. The results are shown in Fig.~\ref{fig:Ac-vs-Cl_gaussian_cl_p2}. As expected initially, in this scenario, the power spectrum captures the anisotropy more effectively than the autocorrelation function. It is particularly evident for small values of $\sigma$, where a single multipole is the most pronounced.

We can conclude that for this family of models, the autocorrelation function proves to be more sensitive than the power spectrum, unless the central mode ($l_0$) is too high.\footnote{ While in the current setup the $p$-value uncertainty bands of the autocorrelation function and the angular power spectrum overlap, it is natural to expect their monotonic shrinking and separation with the growth of event statistics in the set.}. It is interesting to note, that the cumulative power spectrum (Eq.~\ref{eq:cl_sum}) shows systematically lower sensitivity than both the standard power spectrum and the cumulative autocorrelation function. At the same time, this result is not surprising for the particular model, as the accumulation of multipoles only enhances the smearing of the central mode.

%\item \textbf{Turbulent diffusion model.}
\subsection{Turbulent diffusion model}
\label{sec:turb_Cl}
\begin{figure}[!ht]
	\centering
    \includegraphics[width=0.9 \linewidth]{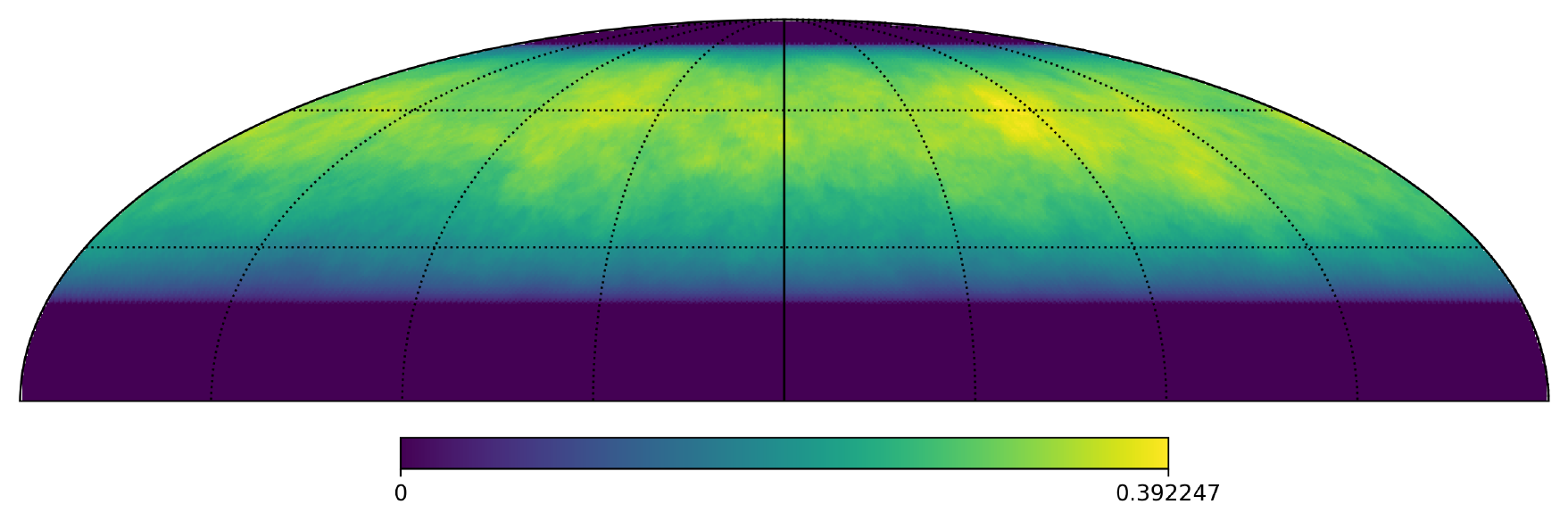}
    \caption{The reference intensity map with a power spectrum following Eq.~\eqref{eq:ahlers} and taking into account the detector exposure.}
    \label{fig:ahlers_map}
\end{figure}

As a physically motivated scenario for small- and medium-scale anisotropy, we adopt the model of CR turbulent diffusion in the Galaxy, proposed in  Ref.~\cite{PhysRevLett.112.021101}. In this model, the scattering of CRs within the local turbulent magnetic field redistributes the large-scale dipole gradient into smaller angular scales. Consequently, this turbulent cascade is described by the following angular power spectrum (see Fig.~\ref{fig:ahlers_map}):
\begin{equation}
C_l = \frac{1}{(2l+1)(l+1)(l+2)}.
\label{eq:ahlers}
\end{equation}
For this model, we set $\epsilon = 0.25$ in Eq.~\eqref{eq:anisotropy}.

This model also strongly favors low multipoles, which makes it comparable to the previous model at medium angular scales. As expected, the median post-trial $p$-value behaves similarly to the model of Sec.~\ref{sec:gaussian_Cl} at low multipoles, with the autocorrelation approach yielding a lower $p$-value (see Table \ref{tab:p_values_comparison}).
\begin{table}[ht]
\centering
\begin{tabular}{l|c|c}
\hline
Method & Median post-trial $p$-value & from the 25th to the 75th percentile \\
\hline
Autocorrelation & 0.0055 & 0.0013 to 0.0545 \\
Binned power spectrum & 0.0618 & 0.0061 to 0.3275 \\
Cumulative power spectrum & 0.2459 & 0.0701 to 0.5289 \\
\hline
\end{tabular}
\caption{Comparison of median post-trial $p$-values for the turbulent diffusion model~\eqref{eq:ahlers}. The intervals indicate the range around the median: from the 25th to the 75th percentile.}
\label{tab:p_values_comparison}
\end{table}

%\item \textbf{Point source.}
\subsection{Single source model}    
\label{sec:single_source}
\begin{figure}[!ht] 
	\centering
    \includegraphics[width=0.9 \linewidth]{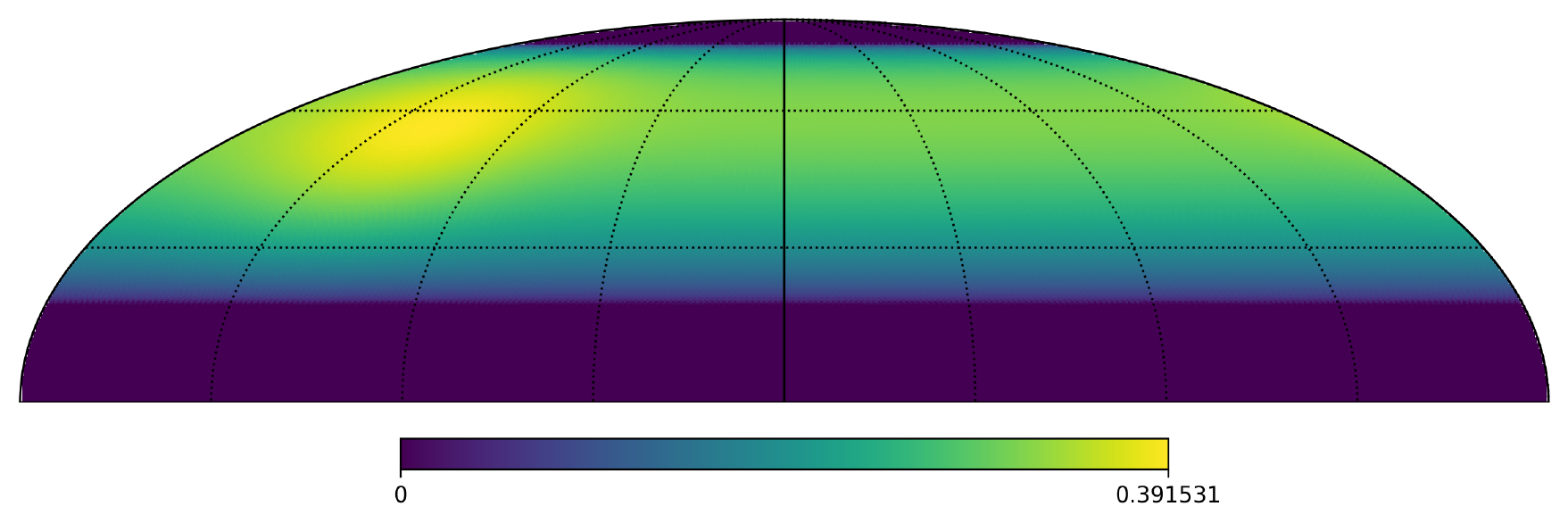}
    \caption{The reference intensity map for the Gaussian source model with $\sigma = 20^\circ,~\alpha = 120^\circ,~\delta = 50^\circ$. The detector exposure is taken into account.}
    \label{fig:gaussian_source}
\end{figure}

The final model considered is the Gaussian source model, representing a single localized source with a Gaussian profile:
\begin{equation}
f(\vec{n}) = \exp\left(-\frac{\psi^2(\vec{n}, \vec{n}_0)}{2\sigma^2}\right),
\label{eq:gauss_source}
\end{equation}
where $\psi(\vec{n}, \vec{n}_0)$ denotes the angular distance between an arbitrary arrival direction $\vec{n}$ and the direction of the source center $\vec{n}_0$, while $\sigma$ characterizes the angular size of the source. For this case, the anisotropy level is set to $\epsilon = 0.2$ in Eq.~\eqref{eq:anisotropy}. This scenario describes another simplified, yet physically motivated model of anisotropy. We assume that a nearby source is located within the detector's field of view, with its angular size determined by the parameter $\sigma$. The position of the source $(\alpha, \delta)$ is fixed: the results for sources partially covered by FoV are similar if we upscale the parameter $\epsilon$. The intensity map for this model is shown in Fig.~\ref{fig:gaussian_source}. 

\begin{figure}[!ht]
    \centering
    \begin{subfigure}[b]{0.8\textwidth}
        \centering
        \includegraphics[width=\textwidth]{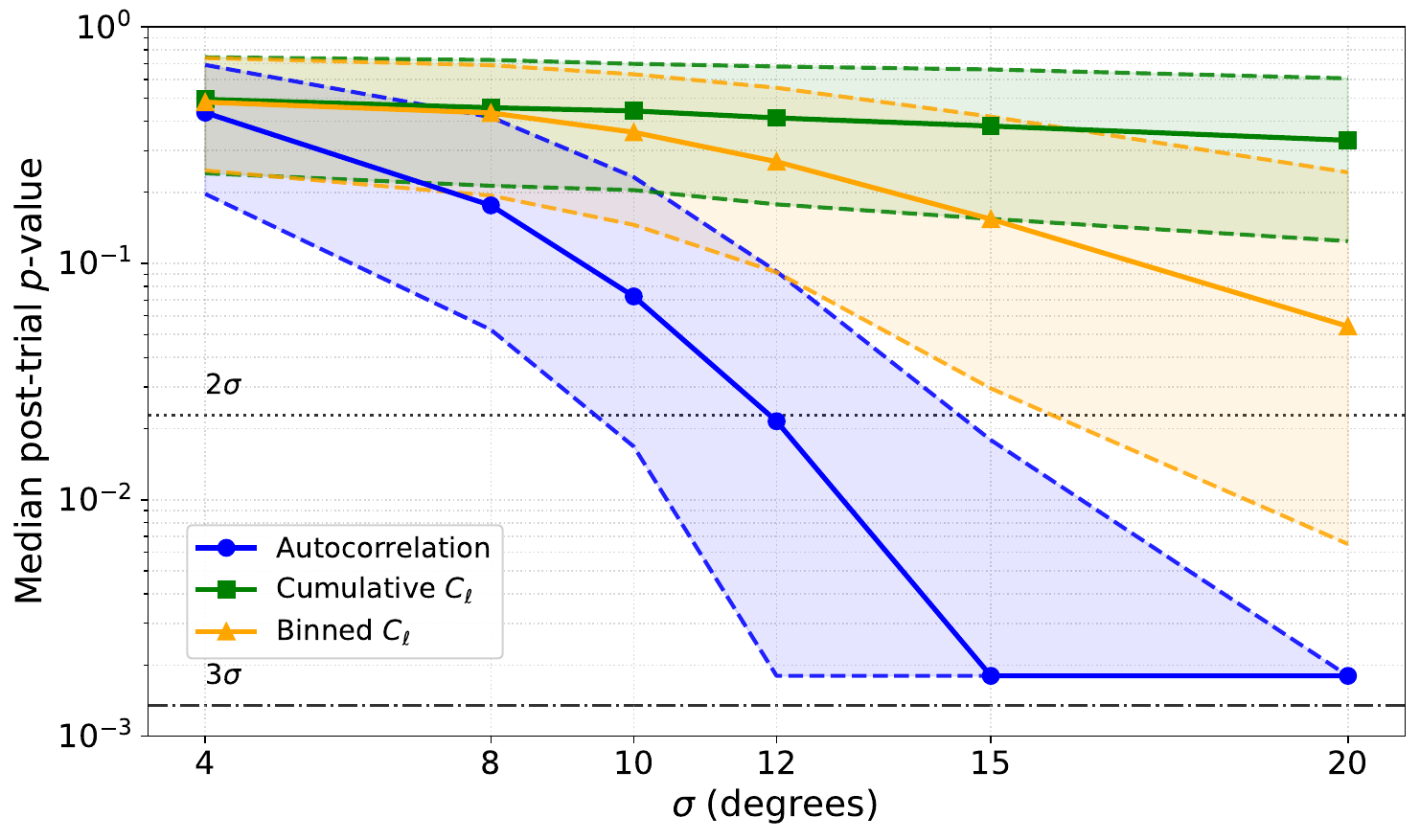}
    \end{subfigure}
    \caption{Comparison between the autocorrelation function and power spectrum for a Gaussian source (Eq.~\ref{eq:gauss_source}) with varying $\sigma$. The shaded bands denote the range around the median: from the 25th to the 75th percentile.}
    \label{fig:Ac-vs-Cl_gaussian_source}
\end{figure}

An analysis of this model across various angular source sizes revealed that the power spectrum is entirely insensitive to this type of anisotropy (see Fig.~\ref{fig:Ac-vs-Cl_gaussian_source}). Furthermore, the signal from this hot spot is distributed across a wide range of multipole modes $l$. One might naively expect this to manifest in the cumulative power spectrum, but its statistical significance is even smaller than that of binned $C_l$.  It is also important to note that the autocorrelation function is significantly less sensitive to features at small angular scales, similar to the case of small-scale anisotropy in the Gaussian power spectrum model (Eq.~\ref{eq:gauss}). A key result here is the high significance of the autocorrelation function observed for $\sigma \gtrsim 10^\circ$.

%\end{enumerate}

In summary, we have analyzed the sensitivity of the autocorrelation function and the angular power spectrum for several simple anisotropy scenarios, including some physically motivated models. A key finding is the comparatively high sensitivity of the autocorrelation function to various types of anisotropy, which highlights the strength of this approach for anisotropy analysis. It is interesting to note that the cumulative power spectrum appears to be less sensitive than both the binned power spectrum and the autocorrelation function for all the anisotropy models considered. This result shows that the expectation of accumulating significance by integrating minor excesses over a wide angular scale (or multipole modes $l$) does not work a priori. In turn, this indicates that the higher sensitivity of the cumulative autocorrelation function relative to the binned angular power spectrum may not be related to this integration effect.

\section{Experiment, data and Monte Carlo}
\label{sec:expMC}

Our aim is to apply the methods discussed in the previous section to the archival data of the KASCADE experiment~\cite{KASCADE:2003swk}. The KASCADE air-shower array operated from 1996 to 2013 at the KIT Campus in Karlsruhe, Germany ($49^\circ$ N, $8.4^\circ$ E, 110~m a.s.l.). The experiment measured extensive air showers~(EAS) in the primary energy range from approximately 500~TeV to 100~PeV. Although KASCADE collected data in several configurations, in this work we focus exclusively on the main KASCADE array. This array consisted of 252 scintillator detectors arranged in a rectangular grid covering an area of $200 \times 200,\mathrm{m}^2$. The outer 192 detectors included a shielding layer to allow for separate detection of the electromagnetic and muon-dominated components of air showers.

The experimental and Monte Carlo data we are using in this study were provided by the KCDC service~\cite{Haungs_2018}. Each event includes time-integrated deposits of electromagnetic and muon EAS components from the KASCADE array stations, as well as reconstructed shower parameters: primary energy ($E$), zenith angle ($\theta$), azimuthal angle ($\phi$), shower core position ($x$, $y$), number of electrons ($N_e$), number of muons ($N_{\mu}$), and shower age ($s$). The values $N_e$, $N_{\mu}$, and $s$ are determined by fitting the lateral distribution function of the particle densities with the modified Nishimura-Kamata-Greisen function.  $E$ is reconstructed with the standard KASCADE algorithms by taking into account both $N_e$ and $N_{\mu}$ corrected for atmospheric attenuation depending on $\theta$. A detailed description of these parameters and the reconstruction procedure can be found in Ref.~\cite{KCDC_manual_2013}. In this work, we use the quality cuts recommended by KASCADE: $x^2+y^2<(91~m)^2$, $\lg N_e>4.8$, $\lg N_\mu>3.6$, and the cut on the shower age set by KCDC: $0.2<s<1.48$. Additionally, we consider zenith angles $\theta < 30^\circ$ and energies $\lg (E/\text{eV})>15.15$. The primary CR direction is reconstructed with an angular resolution of approximately $0.23^\circ$ (68.3\% quantile).
The resolution of energy reconstruction is around $11\%$ in terms of the decimal logarithm of the ratio between the simulated and the reconstructed energies. Both estimates are derived using Monte Carlo simulations based on the QGSJet-II.04 hadronic interaction model, considering only the events that passed the quality cuts within the studied energy range.

In our previous studies, the KASCADE data were classified into five primary mass groups: protons (p), helium (He), carbon (C), silicon (Si) and iron (Fe), using convolutional neural network technique~\cite{Kuznetsov:2023kss, Kuznetsov:2023pvo}. In the present work we use this classification to study the anisotropy of the individual CR mass components. Namely, we apply the methods of anisotropy estimation considered in Sec.~\ref{sec:method} to a sample of KASCADE data. To select the optimal parameters and methods for anisotropy detection, an ``unblind'' dataset is used in this study. It comprises of 10\% of the full dataset, that is 800,487 randomly selected events after all quality cuts. These events were classified by a convolutional neural network trained on the Monte Carlo datasets generated using the QGSJet-II.04 hadronic model (see Ref.~\cite{Kuznetsov:2023pvo} for details).

\begin{figure}[!ht]
    \centering
    \begin{subfigure}[b]{0.48\textwidth}
        \centering
        \includegraphics[width=\textwidth]{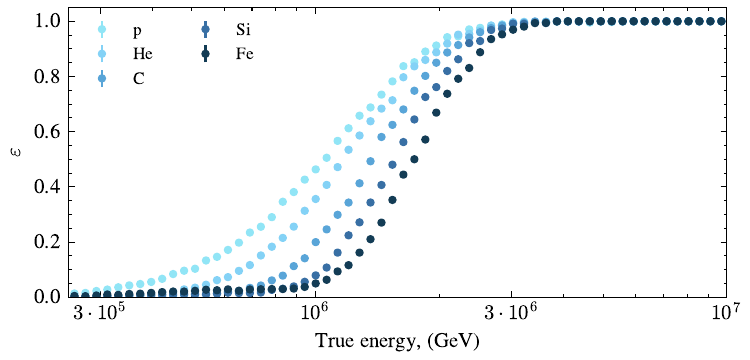}
    \end{subfigure}
    \hfill
    \begin{subfigure}[b]{0.48\textwidth}
        \centering
        \includegraphics[width=\textwidth]{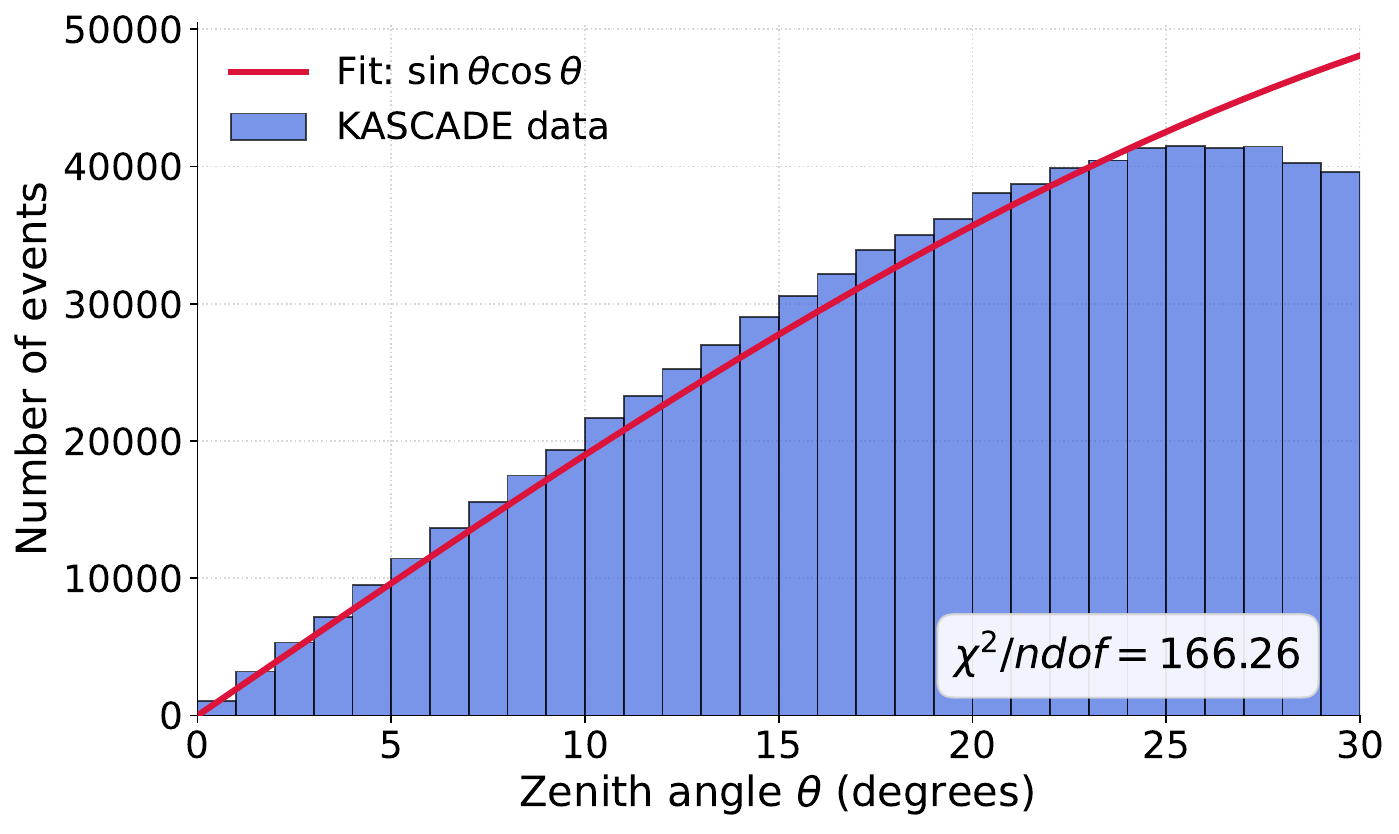}
    \end{subfigure}
    \caption{\textbf{Left panel:} Detection efficiency (after quality cuts) as a function of energy for different types of primary particles. Monte Carlo simulations based on the QGSJet-II.04 hadronic interaction model are used. \textbf{Right panel:} Zenith angle distribution for the all-particle ``unblind'' set at energies $\lg (E/\text{eV}) > 15.15$.}
    \label{fig:eff_and_zenith}
\end{figure}

The efficiency of shower detection after the applied quality cuts depends on both the primary particle type and its energy. An analysis of this efficiency reveals that the detector exposure deviates from a purely geometric form, particularly in the lower energy bins (see Fig.~\ref{fig:eff_and_zenith}). Furthermore, diurnal variations in the event rate are also present in the data. These factors necessitate the use of dedicated methods for reconstructing the actual detector exposure.

\begin{figure}[!ht]
	\centering
    \includegraphics[width=0.7 \linewidth]{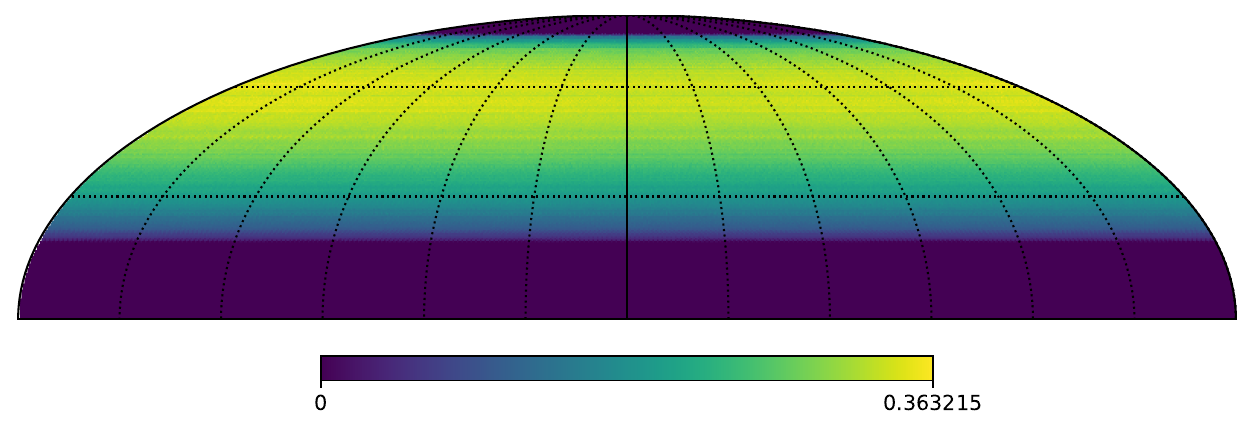}
    \caption{Relative exposure map obtained using the time-scrambling method for protons with energy $\lg(E/\text{eV}) > 15.15$.}
    \label{fig:scrambled_map}
\end{figure}

In this work, the time-scrambling method \cite{ALEXANDREAS1993570} is used to reconstruct the detector exposure. The algorithm creates artificial events by randomly combining the local arrival direction of one event with the arrival time of another from the dataset. Based on this new ``local coordinates -- time'' pair, the equatorial coordinates for the background event are calculated. As a result, the simulated sample preserves the original angular distribution while accurately reflecting the non-uniform distribution of time of the KASCADE data. Repeating this procedure, we generate a background sample containing $N = 5 \cdot 10^8$ events. This sample is then projected onto a sky map and normalized to produce the relative exposure map.  While this averaging effectively accounts for instrumental and diurnal effects, it can also smear out extended physical structures, leading to an overestimation of the background and a consequent underestimation of the anisotropy significance in the data. In this study, we apply a separate scrambling procedure for each particle type and energy bin. An example of the relative exposure map obtained using this method is shown in Fig.~\ref{fig:scrambled_map}.

%\section{Analysis and results}
\section{Application of methods to experimental data}
\label{sec:results}
In this section, we evaluate the anisotropy in the KASCADE data for each particle type and for several open energy bins: $\lg(E / \text{eV}) > \{15.15, 15.25, 15.5, 15.75, 16.0, 16.25, 16.5\}$.
The number of events for each primary particle type and energy threshold is shown in Table~\ref{tab:Nevs}. Note that in the computation of the post-trial $p$-value for the small- and medium-scale analyzes, we penalize the result only for the scan over the angular scale $\psi$ and the multipole $\ell$. The different energy bins and primary particle types are considered independent

\begin{table}[htbp]
\centering
\begin{tabular}{|c|c|c|c|c|c|c|c|}
\hline
Type & \multicolumn{7}{c|}{$\lg(E_{thr}/\mathrm{eV})$} \\
\cline{2-8}
 & 15.15 & 15.25 & 15.50 & 15.75 & 16.00 & 16.25 & 16.50 \\
\hline
All & 800,487 & 555,172 & 191,654 & 61,357 & 19,122 & 6,122 & 2,045 \\
p   & 288,286 & 192,877 & 66,059  & 19,256 & 4,756  & 1,154 & 303  \\
He  & 221,074 & 153,137 & 53,048  & 16,456 & 4,582  & 1,175 & 316  \\
C   & 146,927 & 104,543 & 35,701  & 12,050 & 4,061  & 1,363 & 443  \\
Si  & 86,596  & 63,714  & 22,441  & 8,162  & 3,222  & 1,281 & 497  \\
Fe  & 57,604  & 40,901  & 14,405  & 5,433  & 2,501  & 1,149 & 486  \\
\hline
\end{tabular}
\caption{Number of events by energy threshold and primary particle type.}
\label{tab:Nevs}
\end{table}

\subsection{Large-scale anisotropy}
\label{sec:lsa_results}
\begin{figure}[!ht]
	\centering
    \includegraphics[width=0.7 \linewidth]{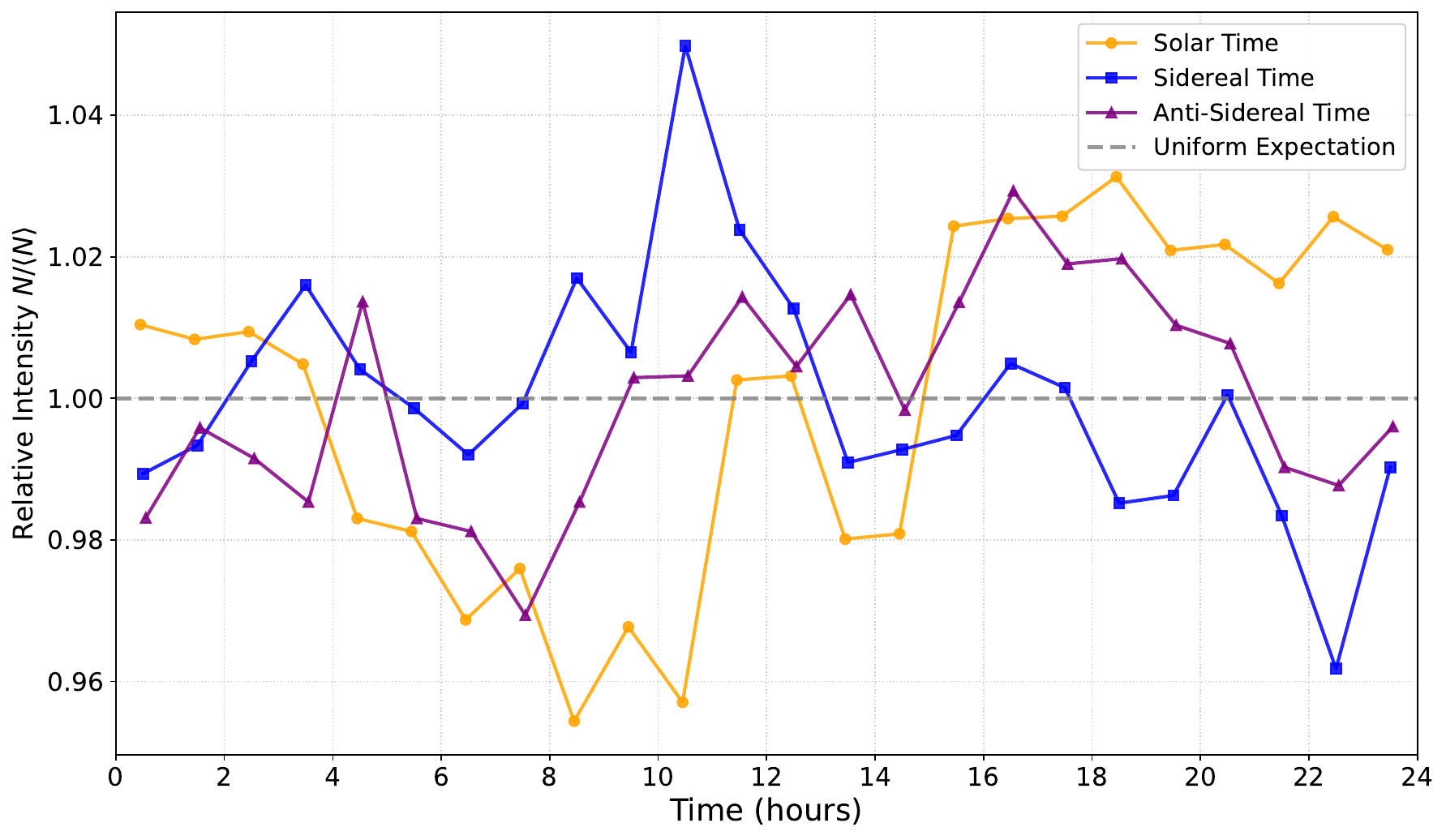}
    \caption{Distributions of solar, sidereal, and anti-sidereal time for proton events with $\lg(E/\text{eV}) > 15.15$.}
    \label{fig:time_distrib_protons}
\end{figure}

We begin with the one-dimensional analysis of large-scale anisotropy using the East-West method~\eqref{eq:eastwest} with respect to solar and anti-sidereal time.
This allows us to test the possible leakage of the dipole component induced by diurnal variations into the physical result evaluated in sidereal time. Figure~\ref{fig:time_distrib_protons} shows the event distributions in solar, sidereal, and anti-sidereal time.
The presence of diurnal variations justifies the use of the East-West method for the large-scale anisotropy analysis. Furthermore, this underscores the necessity to use time-scrambling to reconstruct the detector exposure for the study of small- and medium-scale anisotropy.

\clearpage

\begin{figure}[!ht]
	\centering
    \includegraphics[width=0.7 \linewidth]{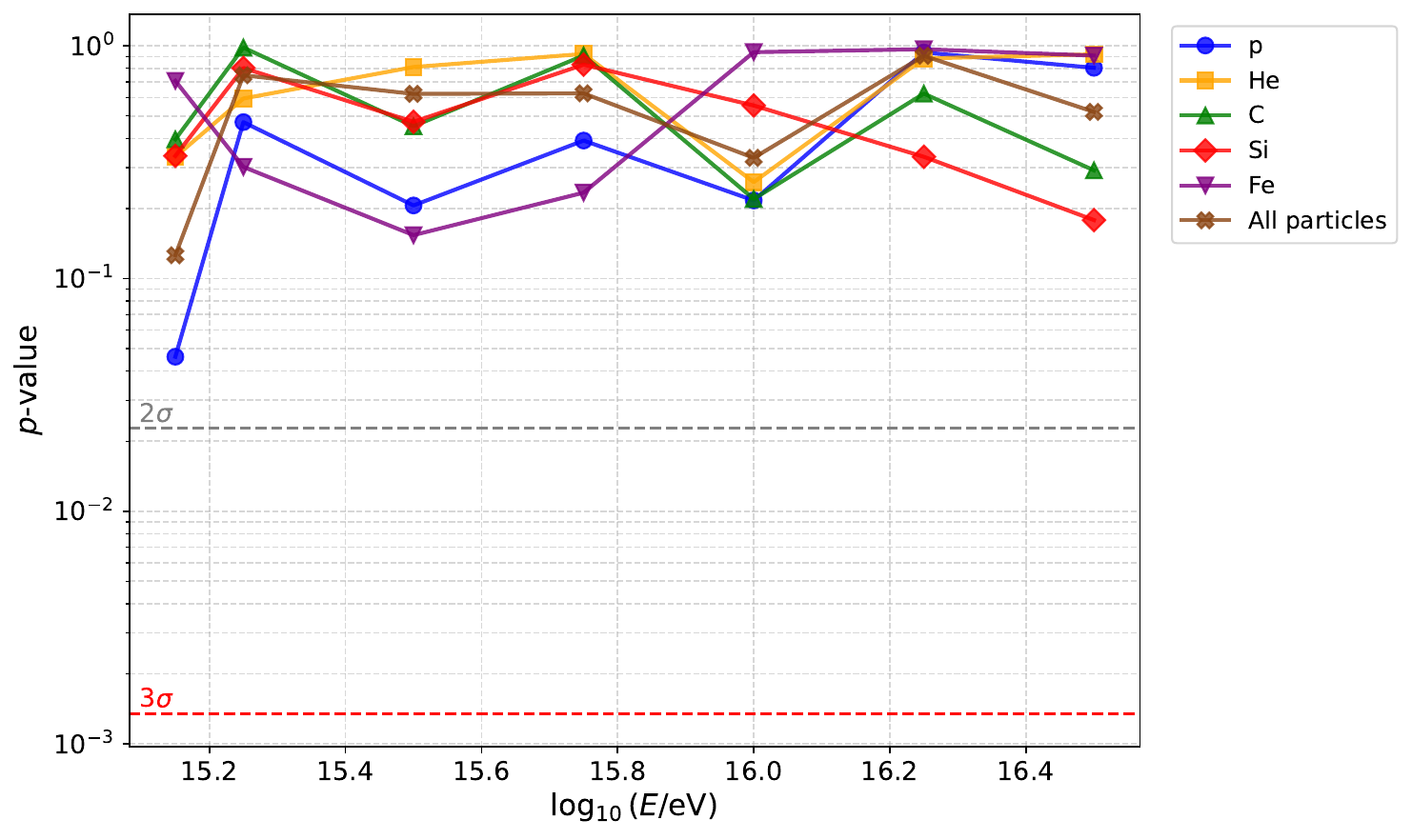}
    \caption{$p$-values of dipole in sidereal time obtained with the East-West method for different primary particle types and energy intervals.}
    \label{fig:res_ew_pvals}
\end{figure}

\begin{figure}[!ht]
	\centering
    \includegraphics[width=0.9 \linewidth]{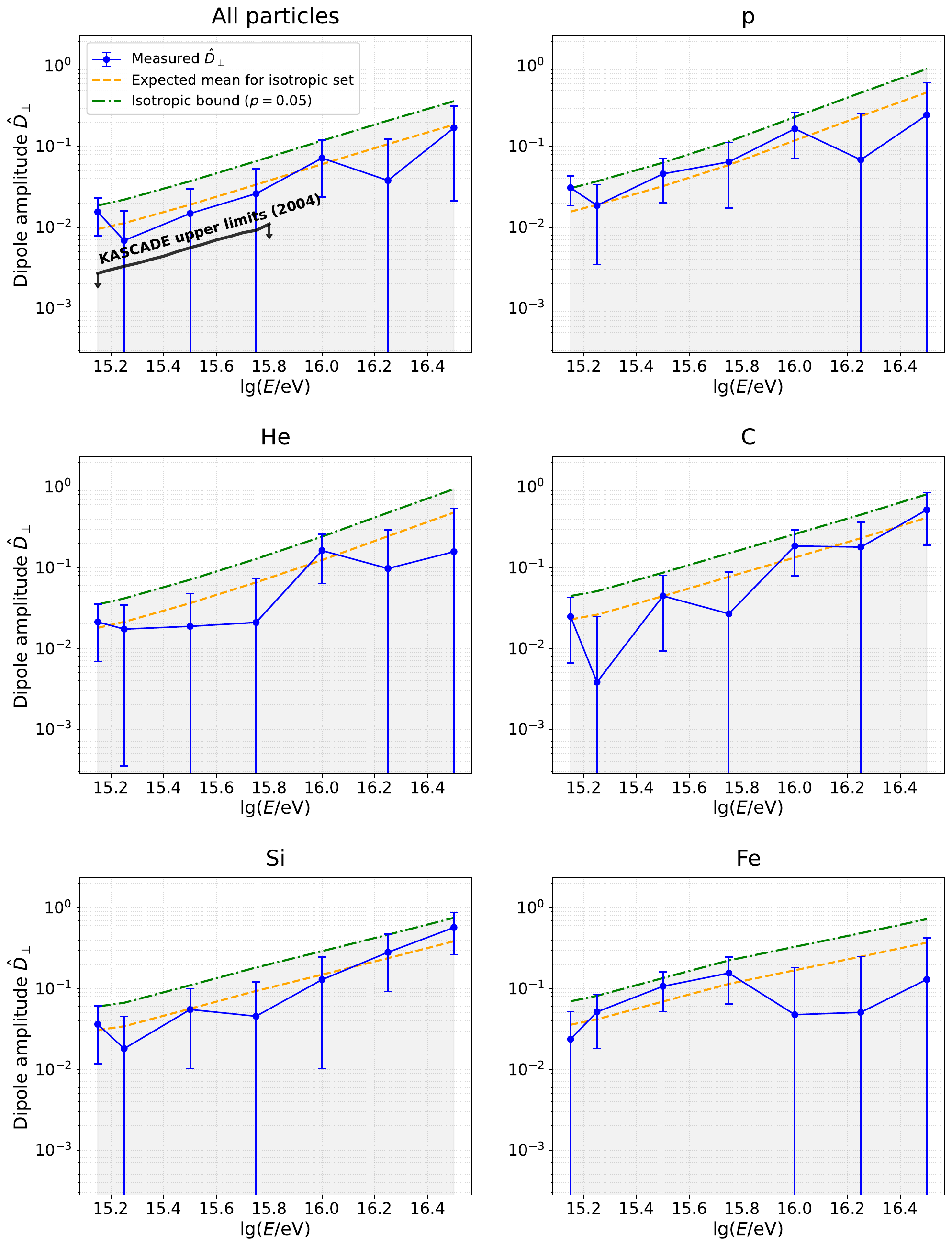}
    \caption{Values of the transverse dipole component derived from the sidereal-time are shown for different primary particle types and energy intervals. The shaded region indicates amplitudes below the dipole amplitude threshold corresponding to a sidereal-time $p$-value of ($p_{sid}=0.05$) under the isotropic hypothesis. For the all-particles sample, the results are compared with the KASCADE upper limits reported in Ref.~\cite{Antoni_2004}.}
    \label{fig:res_ew_dipoles}
\end{figure}

The analysis of the unblinded dataset using the East-West method revealed no statistically significant deviations from isotropy for any primary particle species across all considered energy intervals (see Figs.~\ref{fig:res_ew_dipoles}, \ref{fig:res_ew_pvals})~\footnote{Note that a one-sided conversion of the p-value to significance is used in Fig.~\ref{fig:res_ew_pvals} and similar subsequent figures because in this analysis we are interested only in excesses but not in deficits with respect to isotropy.} The lowest $p$-value obtained in sidereal time was $p_{sid} = 0.046$ for protons at $\lg(E/\text{eV}) > 15.15$. It is noteworthy that our results for the all-particle dataset differ from the KASCADE upper limits reported in Ref.~\cite{Antoni_2004}. This discrepancy is explained by the size of the analyzed statistics: the KASCADE upper limits were derived from approximately $2 \times 10^7$ events, whereas our lowest energy bin contains only $8 \times 10^5$ events. This difference inevitably leads to larger statistical fluctuations in the isotropic background. A similar situation can be observed in Ref.~\cite{Kostunin:2021eyp}, where an analysis of a comparable number of events yielded results consistent with isotropy, with an amplitude approximately equal to ours.

\clearpage

\subsection{Autocorrelation}

\begin{figure}[!ht]
	\centering
    \includegraphics[width=0.7 \linewidth]{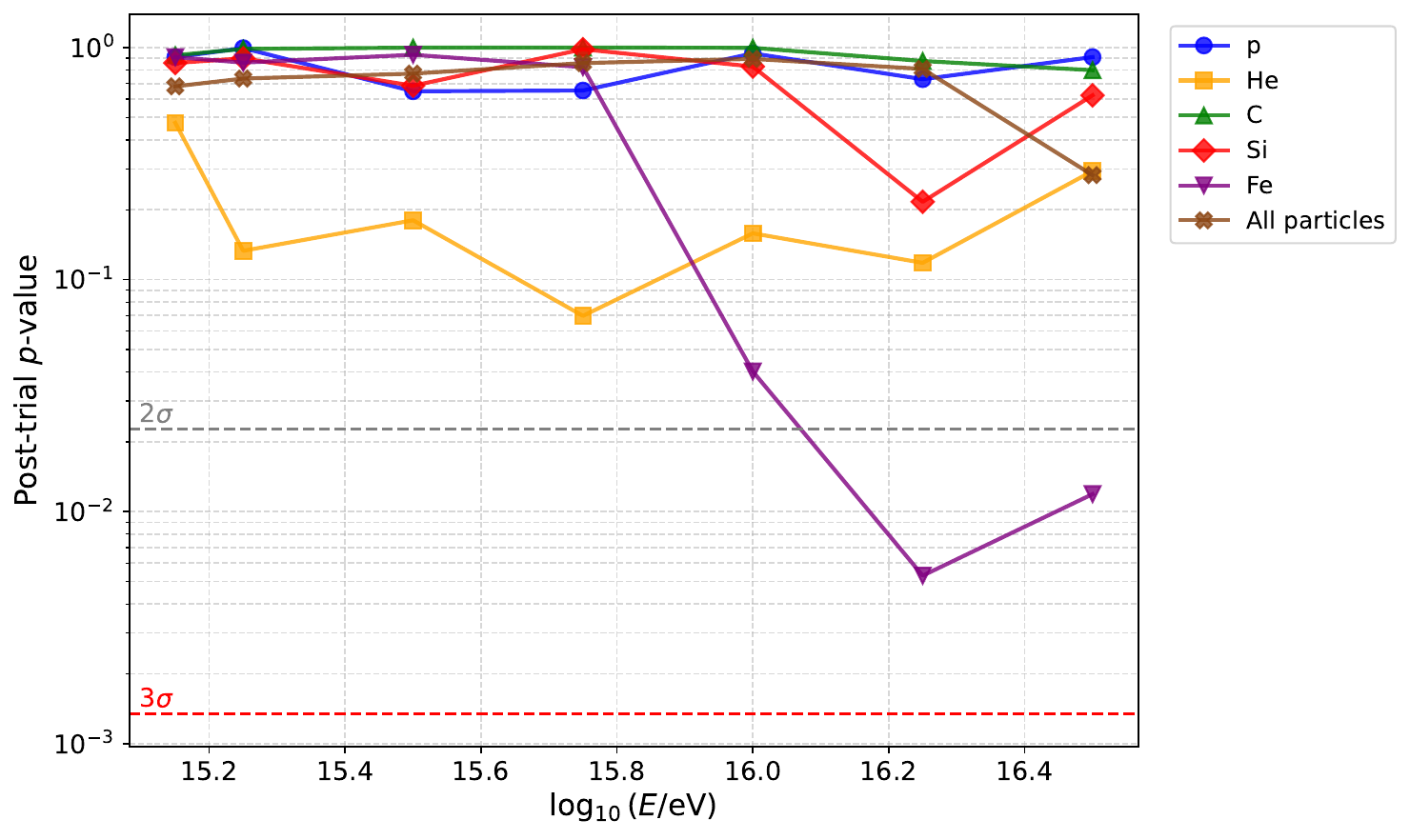}
    \caption{The post-trial $p$-values obtained using the autocorrelation method for various primary particle species and energy intervals.}
    \label{fig:res_autocorr_pvals}
\end{figure}

\begin{figure}[!ht]
    \centering
    \begin{subfigure}[b]{0.48\textwidth}
        \centering
        \includegraphics[width=\textwidth]{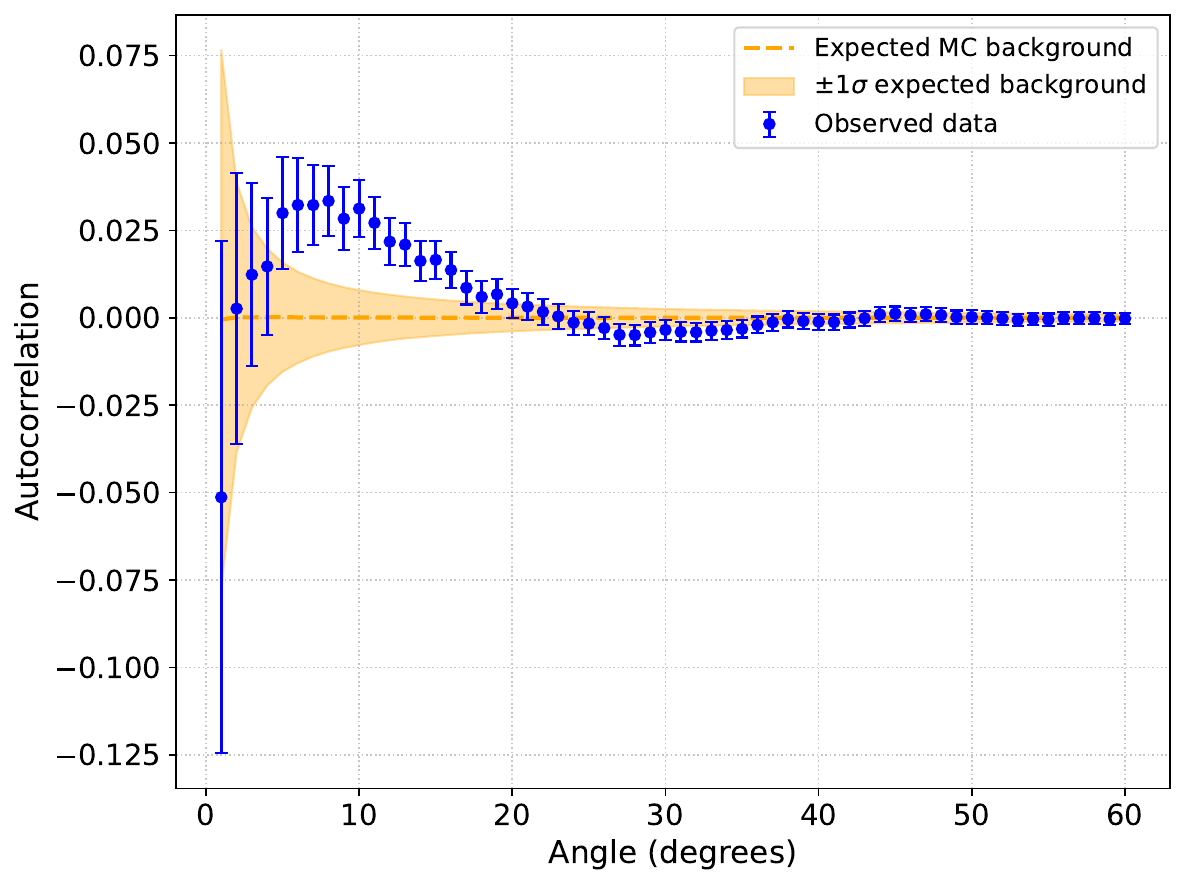}
        \caption{}
        \label{fig:res_autocorr_w_fe}
    \end{subfigure}
    \hfill
    \begin{subfigure}[b]{0.48\textwidth}
        \centering
        \includegraphics[width=\textwidth]{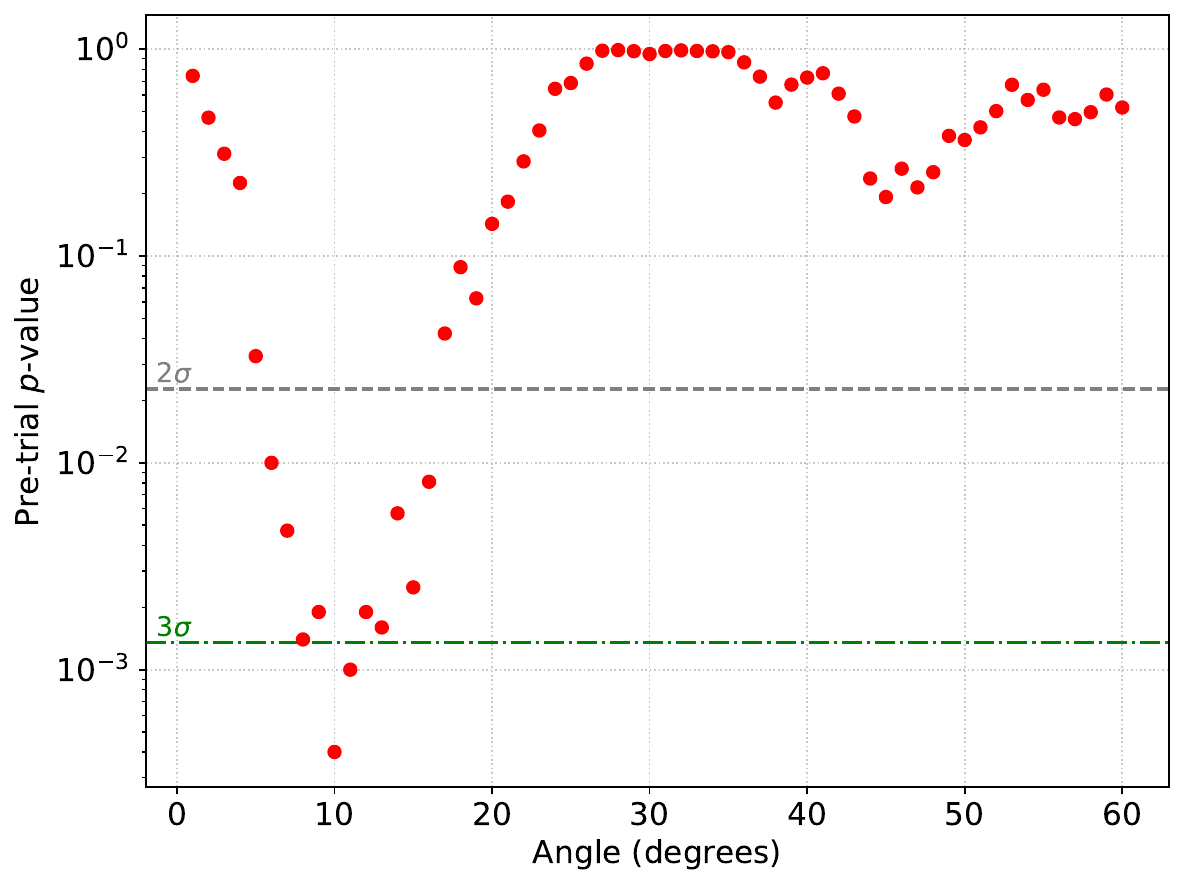}
        \caption{}
        \label{fig:res_autocorr_pvals_fe}
    \end{subfigure}
    \caption{Results of the autocorrelation analysis for iron at energies $\lg(E/\text{eV}) > 16.25$. \textbf{Left panel:} the autocorrelation function. \textbf{Right panel:} the corresponding pre-trial $p$-values.}
    \label{fig:res_autocorr_fe_combined}
\end{figure}

We computed the cumulative autocorrelation function using the Landy-Szalay estimator Eq.~\eqref{eq:autocorr}. The results for various primaries and energies are shown in Fig.~\ref{fig:res_autocorr_pvals} in terms of post-trail $p$-values. One can see, that no significant deviations from isotropy is observed, except the case of iron nuclei at energies $\lg(E/\text{eV}) > 16.25$ that yields a minimum post-trial $p$-value of $0.0053$.
The respective distribution of autocorrelation has a peak value around $10^\circ$ with the pre-trial $p$-value equal to $4 \times 10^{-4}$; see Fig.~\ref{fig:res_autocorr_fe_combined}. Note that for the all-particle set, because of the algorithmic complexity of computing the autocorrelation function and the large number of events at energies $\lg(E/\text{eV}) > \{15.15, 15.25, 15.5\}$, it was calculated and analyzed only up to $20^\circ$ angular separation.

\subsection{Angular power spectrum}

\begin{figure}[!ht]
	\centering
    \includegraphics[width=0.7 \linewidth]{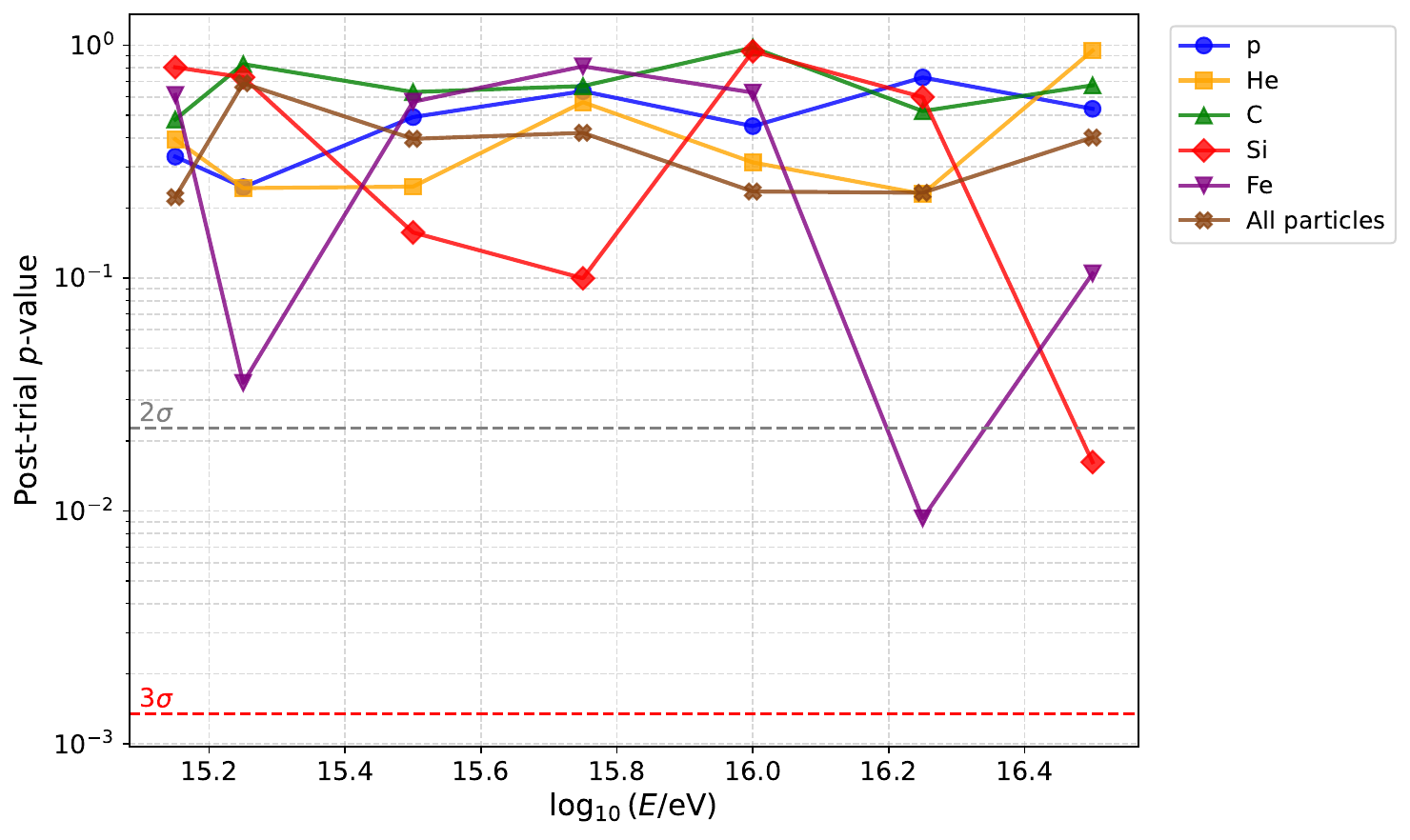}
    \caption{The post-trial $p$-values obtained using the power spectrum method for various primary particle species and energy intervals.}
    \label{fig:res_ps_pvals}
\end{figure}

\begin{figure}[!ht]
    \centering
    \begin{subfigure}[b]{0.48\textwidth}
        \centering
        \includegraphics[width=\textwidth]{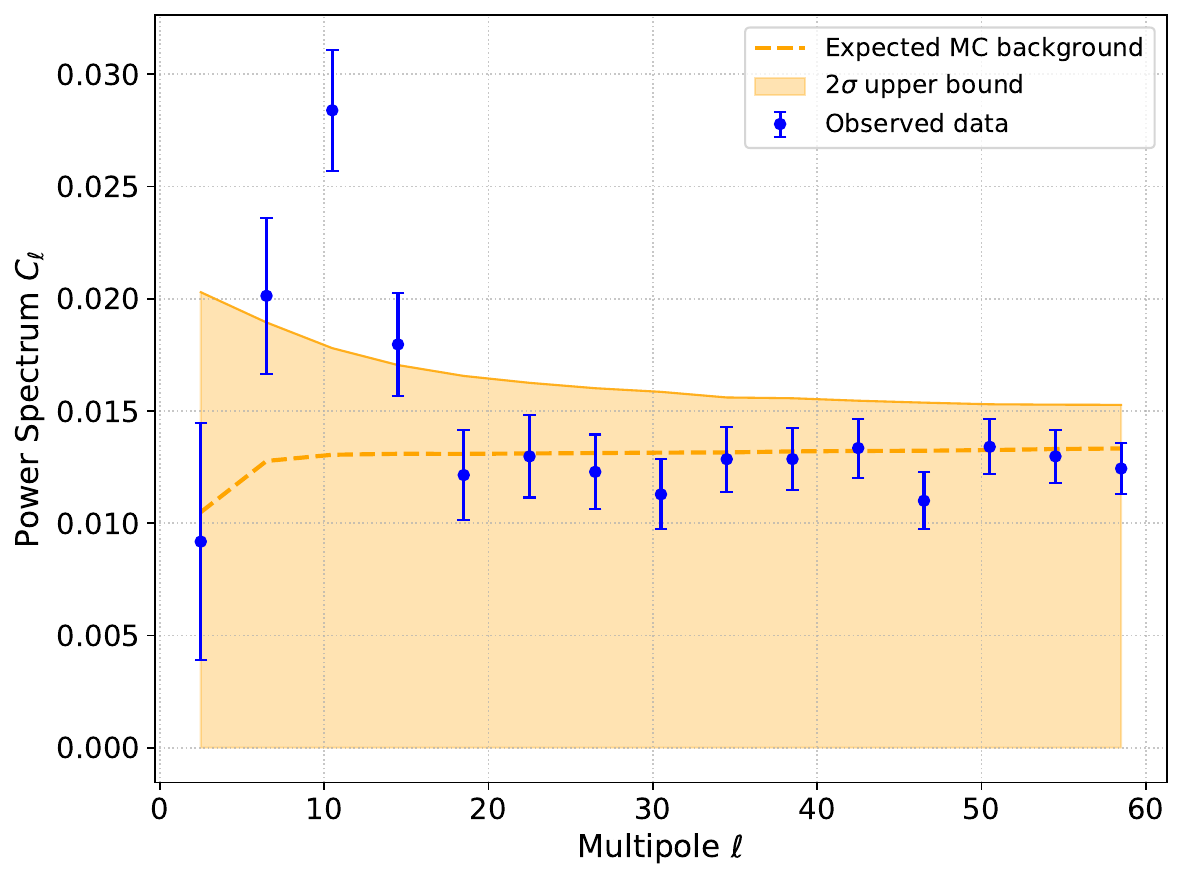}
        \caption{}
        \label{fig:res_ps_cl_fe}
    \end{subfigure}
    \hfill
    \begin{subfigure}[b]{0.48\textwidth}
        \centering
        \includegraphics[width=\textwidth]{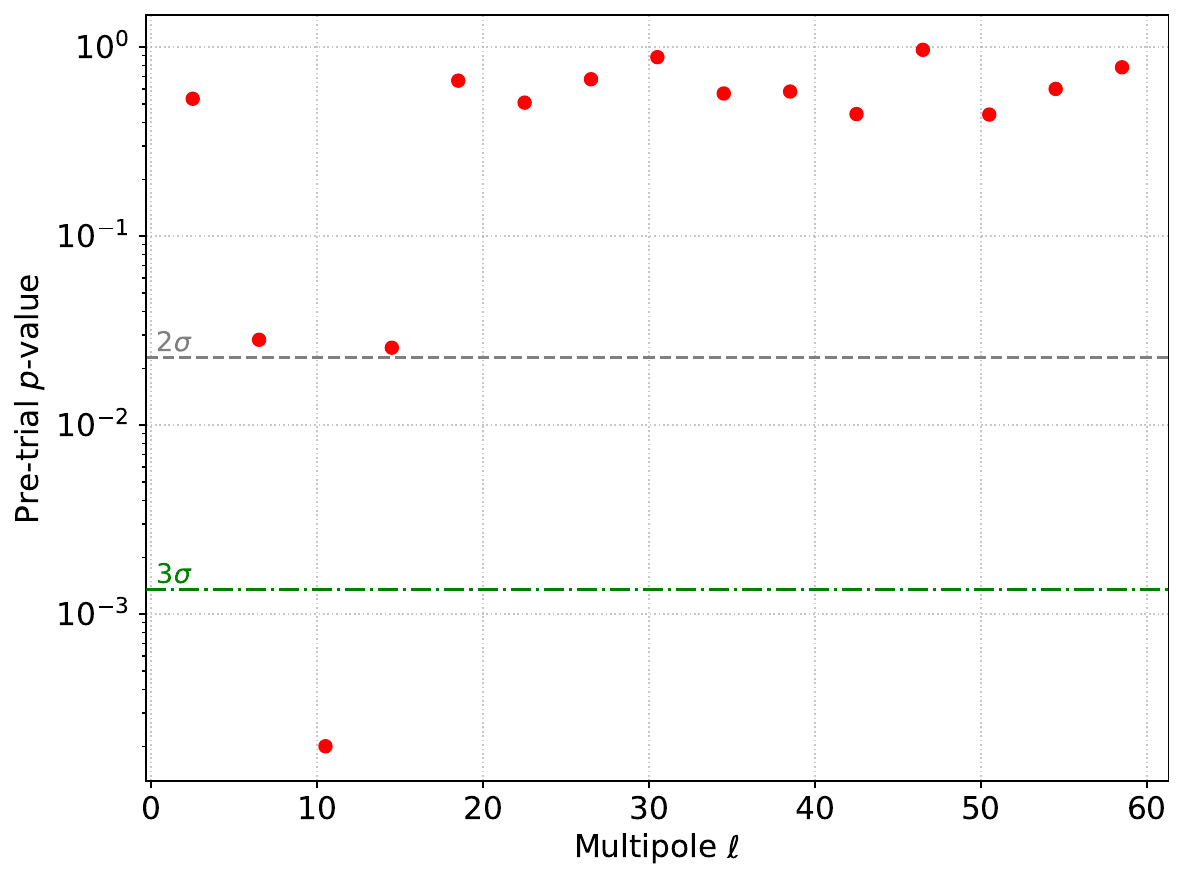}
        \caption{}
        \label{fig:res_ps_pvals_fe}
    \end{subfigure}
    \caption{Results of the power spectrum analysis for iron at energies $\lg(E/\text{eV}) > 16.25$. \textbf{Left panel:} the power spectrum. \textbf{Right panel:} the corresponding pre-trial $p$-values.}
    \label{fig:res_ps_fe_combined}
\end{figure}

We calculate the angular power spectrum using the standard binned method~\eqref{eq:std_Cl}; the cumulative form~\eqref{eq:cl_sum} is not used, as it proved to be insensitive in Sec.~\ref{sec:method}. Similarly to the autocorrelation method, no significant deviations from isotropy were observed for the majority of particle types and in most energy intervals. The exceptions are Fe at the energies $\lg(E/\text{eV}) > 16.25$ and $\lg(E/\text{eV}) > 16.5$, and Si at $\lg(E/\text{eV}) > 16.5$. The respective post-trial $p$-values are: $0.0093$, $0.0119$, and $0.0162$; see Fig.~\ref{fig:res_ps_pvals}. The angular power spectrum for Fe at energies $\lg(E/\text{eV}) > 16.25$ and the respective pre-trial $p$-values are shown in Fig.~\ref{fig:res_ps_fe_combined}. The minimum $p$-value of $2 \times 10^{-4}$ is reached around $l=12$.

\clearpage

%%%%%%%%%%%%%%%%%%%%%%%%%%%%%%%%%%%%%%%%%%%%%%%%%%%%%%%%%
\section{Discussion and conclusion}
\label{sec:conclusion}
In this work, we examined two primary methods for anisotropy analysis: the autocorrelation function and the angular power spectrum. We compared their sensitivities to medium- and small-scale anisotropy for an experiment with a limited field of view and geometric exposure. We found that the autocorrelation function method has a systematically higher sensitivity than the angular power spectrum method for most of the anisotropy models considered. At the same time, if there is a real detectable anisotropy in the dataset, the application of both of these methods would facilitate a better understanding. We also found that for experiments with a limited field of view, the Landy-Szalay estimator for the autocorrelation function has a significantly higher sensitivity to medium scale anisotropy compared to the standard estimator. 

\begin{figure}[!ht]
	\centering
    \includegraphics[width=0.8 \linewidth]{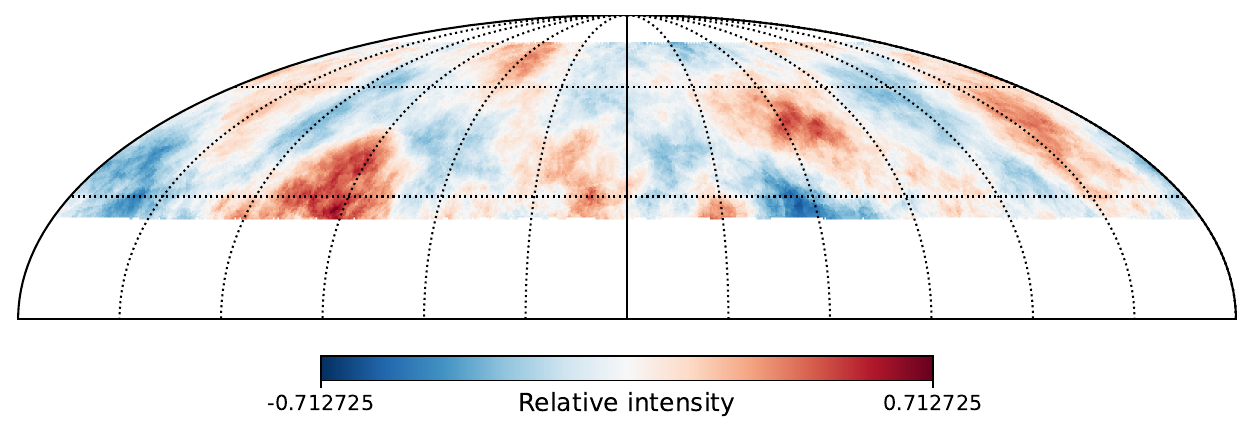}
    \caption{Relative intensity map ($12^\circ$ smoothing) for the Fe mass group, $\lg(E/\text{eV}) > 16.25$.}
    \label{fig:intensity_map}
\end{figure}

As a test of the considered method of anisotropy detection, we applied them to the "unblind" set of the KASCADE experiment data, comprising 10\% of its full dataset. The anisotropy was searched for across several energy intervals and five mass groups of primary CRs (p, He, C, Si and Fe), classified by the neural network method in our previous studies~\cite{Kuznetsov:2023kss, Kuznetsov:2023pvo}. The search for anisotropy in the respective all-particle sample was also performed. Indications of anisotropy with $> 2.5 \sigma$ post-trail significances (one-sided) were found for the Fe mass group at energies $\lg(E/\text{eV}) > 16.25$ in both the angular power spectrum and the autocorrelation function at angular scales of approximately $10^\circ$ (see also the relative intensity map, Fig.~\ref{fig:intensity_map}). Based on our evaluation of the sensitivity of these two methods, this could indicate the presence of an underlying regular structure at this scale. For the dipole anisotropy search using the East-West method, no significant deviation from isotropy was detected.

The conclusive searches for anisotropies and a robust interpretation of its results would be given with the blind set comprising the remaining $90\%$ of the KASCADE data. However, such a study would require huge computational resources or a significant change in computational techniques, and will be given in our future studies.

\acknowledgments
We thank Grigory Rubtsov, Sergey Troitsky, Ivan Kharuk, Petr Tinyakov, Dmitry Chernyshov and Federico Urban for fruitful discussions. We are also grateful to Dmitry Kostunin for the initial idea of this study, and to Vladimir Sotnikov for managing the preprocessed data used in this analysis. This work was supported by the Russian Science Foundation grant \mbox 25-12-00111. The work of E. Shinkevich was supported by the Foundation for the Advancement of Theoretical Physics and Mathematics ”BASIS”.

\newpage

% Bibliography
\bibliographystyle{JHEP}
\bibliography{refs.bib}

@article{Ahlers:2019gdc,
    author = "Ahlers, Markus",
    title = "{Large- and Medium-Scale Anisotropies in the Arrival Directions of Cosmic Rays observed with KASCADE-Grande}",
    eprint = "1909.09222",
    archivePrefix = "arXiv",
    primaryClass = "astro-ph.HE",
    doi = "10.3847/2041-8213/ab552f",
    journal = "Astrophys. J. Lett.",
    volume = "886",
    number = "1",
    pages = "L18",
    year = "2019"
}

@article{Gao:2023jlz,
    author = "Gao, Wei and He, Huihai and Lv, Hongkui and Cui, Shuwang and Zhang, Weiyan",
    collaboration = "LHAASO",
    title = "{The large-scale anisotropy of cosmic rays based on LHAASO-KM2A}",
    doi = "10.22323/1.444.0478",
    journal = "PoS",
    volume = "ICRC2023",
    pages = "478",
    year = "2023"
}

@article{ARGO-YBJ:2018zoa,
    author = "Bartoli, B. and others",
    collaboration = "ARGO-YBJ",
    title = "{Galactic Cosmic-Ray Anisotropy in the Northern Hemisphere from the ARGO-YBJ Experiment during 2008{\textendash}2012}",
    eprint = "1805.08980",
    archivePrefix = "arXiv",
    primaryClass = "astro-ph.HE",
    doi = "10.3847/1538-4357/aac6cc",
    journal = "Astrophys. J.",
    volume = "861",
    number = "2",
    pages = "93",
    year = "2018"
}

@article{IceCube:2024pnx,
    author = "Abbasi, R. and others",
    collaboration = "IceCube",
    title = "{Observation of Cosmic-Ray Anisotropy in the Southern Hemisphere with 12 yr of Data Collected by the IceCube Neutrino Observatory}",
    eprint = "2412.05046",
    archivePrefix = "arXiv",
    primaryClass = "astro-ph.HE",
    doi = "10.3847/1538-4357/adb1de",
    journal = "Astrophys. J.",
    volume = "981",
    number = "2",
    pages = "182",
    year = "2025"
}

@article{IceCube:2012vve,
    author = "Aartsen, M. G. and others",
    collaboration = "IceCube",
    title = "{Observation of Cosmic Ray Anisotropy with the IceTop Air Shower Array}",
    eprint = "1210.5278",
    archivePrefix = "arXiv",
    primaryClass = "astro-ph.HE",
    doi = "10.1088/0004-637X/765/1/55",
    journal = "Astrophys. J.",
    volume = "765",
    pages = "55",
    year = "2013"
}

@article{Apel:2019afz,
    author = "Apel, W. D. and others",
    title = "{Search for Large-scale Anisotropy in the Arrival Direction of Cosmic Rays with KASCADE-Grande}",
    doi = "10.3847/1538-4357/aaf1ca",
    journal = "Astrophys. J.",
    volume = "870",
    number = "2",
    pages = "91",
    year = "2019"
}

@article{Amenomori:2017jbv,
    author = "Amenomori, M.",
    collaboration = "Tibet AS-gamma",
    title = "{Northern sky Galactic Cosmic Ray anisotropy between 10-1000 TeV with the Tibet Air Shower Array}",
    eprint = "1701.07144",
    archivePrefix = "arXiv",
    primaryClass = "astro-ph.HE",
    doi = "10.3847/1538-4357/836/2/153",
    journal = "Astrophys. J.",
    volume = "836",
    number = "2",
    pages = "153",
    year = "2017"
}

@article{EAS-TOP:2009nld,
    author = "Aglietta, M. and others",
    collaboration = "EAS-TOP",
    title = "{Evolution of the cosmic ray anisotropy above 10{\textasciicircum}{14} eV}",
    eprint = "0901.2740",
    archivePrefix = "arXiv",
    primaryClass = "astro-ph.HE",
    doi = "10.1088/0004-637X/692/2/L130",
    journal = "Astrophys. J. Lett.",
    volume = "692",
    pages = "L130--L133",
    year = "2009"
}

@article{IceCube:2016biq,
    author = "Aartsen, M. G. and others",
    collaboration = "IceCube",
    title = "{Anisotropy in Cosmic-ray Arrival Directions in the Southern Hemisphere Based on six Years of Data From the Icecube Detector}",
    eprint = "1603.01227",
    archivePrefix = "arXiv",
    primaryClass = "astro-ph.HE",
    doi = "10.3847/0004-637X/826/2/220",
    journal = "Astrophys. J.",
    volume = "826",
    number = "2",
    pages = "220",
    year = "2016"
}

@article{Kachelriess:2019oqu,
    author = "Kachelriess, M. and Semikoz, D. V.",
    title = "{Cosmic Ray Models}",
    eprint = "1904.08160",
    archivePrefix = "arXiv",
    primaryClass = "astro-ph.HE",
    doi = "10.1016/j.ppnp.2019.07.002",
    journal = "Prog. Part. Nucl. Phys.",
    volume = "109",
    pages = "103710",
    year = "2019"
}

@ARTICLE{1961PThPS..20....1G,
       author = {{Ginzburg}, V.~L. and {Syrovatsky}, S.~I.},
        title = "{Origin of Cosmic Rays}",
      journal = {Progress of Theoretical Physics Supplement},
         year = 1961,
        month = jan,
       volume = {20},
        pages = {1-83},
          doi = {10.1143/PTPS.20.1},
       adsurl = {https://ui.adsabs.harvard.edu/abs/1961PThPS..20....1G}
}

@article{Hillas:2005cs,
    author = "Hillas, A. M.",
    title = "{Can diffusive shock acceleration in supernova remnants account for high-energy galactic cosmic rays?}",
    doi = "10.1088/0954-3899/31/5/R02",
    journal = "J. Phys. G",
    volume = "31",
    pages = "R95--R131",
    year = "2005"
}

@article{Sommers:2000us,
    author = "Sommers, P.",
    title = "{Cosmic ray anisotropy analysis with a full-sky observatory}",
    eprint = "astro-ph/0004016",
    archivePrefix = "arXiv",
    doi = "10.1016/S0927-6505(00)00130-4",
    journal = "Astropart. Phys.",
    volume = "14",
    pages = "271--286",
    year = "2001"
}

@article{Kuznetsov:2023pvo,
    author = "Kuznetsov, M. Yu. and Petrov, N. A. and Plokhikh, I. A. and Sotnikov, V. V.",
    title = "{Energy spectra of elemental groups of cosmic rays with~the KASCADE experiment data and machine learning}",
    eprint = "2312.08279",
    archivePrefix = "arXiv",
    primaryClass = "astro-ph.HE",
    doi = "10.1088/1475-7516/2024/05/125",
    journal = "JCAP",
    volume = "05",
    pages = "125",
    year = "2024"
}

@article{KASCADE:2003swk,
    author = "Antoni, T and others",
    collaboration = "KASCADE",
    title = "{The Cosmic ray experiment KASCADE}",
    doi = "10.1016/S0168-9002(03)02076-X",
    journal = "Nucl. Instrum. Meth. A",
    volume = "513",
    pages = "490--510",
    year = "2003"
}

@article{PierreAuger:2022axr,
    author = "Abreu, Pedro and others",
    collaboration = "Pierre Auger",
    title = "{Arrival Directions of Cosmic Rays above 32 EeV from Phase One of the Pierre Auger Observatory}",
    eprint = "2206.13492",
    archivePrefix = "arXiv",
    primaryClass = "astro-ph.HE",
    reportNumber = "FERMILAB-PUB-22-491-AD-PPD-SCD-TD",
    doi = "10.3847/1538-4357/ac7d4e",
    journal = "Astrophys. J.",
    volume = "935",
    number = "2",
    pages = "170",
    year = "2022"
}

@article{Haungs_2018,
   title={The KASCADE Cosmic-ray Data Centre KCDC: granting open access to astroparticle physics research data},
   volume={78},
   ISSN={1434-6052},
   url={http://dx.doi.org/10.1140/epjc/s10052-018-6221-2},
   DOI={10.1140/epjc/s10052-018-6221-2},
   number={9},
   journal={The European Physical Journal C},
   publisher={Springer Science and Business Media LLC},
   author={Haungs, A. and Kang, D. and Schoo, S. and Wochele, D. and Wochele, J. and Apel, W. D. and Arteaga-Velázquez, J. C. and Bekk, K. and Bertaina, M. and Blümer, J. and Bozdog, H. and Brancus, I. M. and Cantoni, E. and Chiavassa, A. and Cossavella, F. and Daumiller, K. and de Souza, V. and Di Pierro, F. and Doll, P. and Engel, R. and Fuchs, B. and Fuhrmann, D. and Gherghel-Lascu, A. and Gils, H. J. and Glasstetter, R. and Grupen, C. and Heck, D. and Hörandel, J. R. and Huege, T. and Kampert, K. H. and Klages, H. O. and Link, K. and Łuczak, P. and Mathes, H. J. and Mayer, H. J. and Milke, J. and Mitrica, B. and Morello, C. and Oehlschläger, J. and Ostapchenko, S. and Petcu, M. and Pierog, T. and Rebel, H. and Roth, M. and Schieler, H. and Schröder, F. G. and Sima, O. and Toma, G. and Trinchero, G. C. and Ulrich, H. and Weindl, A. and Zabierowski, J.},
   year={2018},
   month=sep 
}

@techreport{KCDC_manual_2013,
  title        = {KCDC User Manual: Open Access Solution for the KASCADE},
  author       = {Wochele, J. and Kang, D. and Wochele, D. and Haungs, A. and Schoo, S.},
  institution  = {KASCADE Cosmic-ray Data Centre (KCDC)},
  year         = {2013},
  url          = {https://doi.org/10.17616/R3TS4P}
}

@BOOK{1980lssu.book.....P,
       author = {{Peebles}, P.~J.~E.},
        title = "{The large-scale structure of the universe}",
         year = 1980,
       adsurl = {https://ui.adsabs.harvard.edu/abs/1980lssu.book.....P}
}

@ARTICLE{1993ApJ...412...64L,
       author = {{Landy}, Stephen D. and {Szalay}, Alexander S.},
        title = "{Bias and Variance of Angular Correlation Functions}",
      journal = "Astrophys. J.",
         year = 1993,
        month = jul,
       volume = {412},
        pages = {64},
          doi = {10.1086/172900},
       adsurl = {https://ui.adsabs.harvard.edu/abs/1993ApJ...412...64L}
}

@ARTICLE{2004MNRAS.350..914C,
       author = {{Chon}, Gayoung and {Challinor}, Anthony and {Prunet}, Simon and {Hivon}, Eric and {Szapudi}, Istv{\'a}n},
        title = "{Fast estimation of polarization power spectra using correlation functions}",
      journal = "Mon. Not. Roy. Astron. Soc.",
         year = 2004,
        month = may,
       volume = {350},
       number = {3},
        pages = {914-926},
          doi = {10.1111/j.1365-2966.2004.07737.x},
archivePrefix = {arXiv},
       eprint = {astro-ph/0303414},
 primaryClass = {astro-ph},
       adsurl = {https://ui.adsabs.harvard.edu/abs/2004MNRAS.350..914C}
}

@ARTICLE{2001ApJ...548L.115S,
       author = {{Szapudi}, Istv{\'a}n and {Prunet}, Simon and {Pogosyan}, Dmitry and {Szalay}, Alexander S. and {Bond}, J. Richard},
        title = "{Fast Cosmic Microwave Background Analyses via Correlation Functions}",
      journal = "Astrophys. J. Lett.",
         year = 2001,
        month = feb,
       volume = {548},
       number = {2},
        pages = {L115-L118},
          doi = {10.1086/319105},
       adsurl = {https://ui.adsabs.harvard.edu/abs/2001ApJ...548L.115S}
}

@article{Bonino:2011nx,
    author = "Bonino, R. and Alekseenko, V. V. and Deligny, O. and Ghia, P. L. and Grigat, M. and Letessier-Selvon, A. and Lyberis, H. and Mollerach, S. and Over, S. and Roulet, E.",
    title = "{The East-West method: an exposure-independent method to search for large scale anisotropies of cosmic rays}",
    eprint = "1106.2651",
    archivePrefix = "arXiv",
    primaryClass = "astro-ph.IM",
    doi = "10.1088/0004-637X/738/1/67",
    journal = "Astrophys. J.",
    volume = "738",
    pages = "67",
    year = "2011"
}

@article{PhysRevLett.112.021101,
  title = {Anomalous Anisotropies of Cosmic Rays from Turbulent Magnetic Fields},
  author = {Ahlers, Markus},
  journal = {Phys. Rev. Lett.},
  volume = {112},
  issue = {2},
  pages = {021101},
  numpages = {5},
  year = {2014},
  month = {Jan},
  publisher = {American Physical Society},
  doi = {10.1103/PhysRevLett.112.021101},
  url = {https://link.aps.org/doi/10.1103/PhysRevLett.112.021101}
}

@article{Linsley:1975kp,
    author = "Linsley, John",
    title = "{Fluctuation effects on directional data}",
    doi = "10.1103/PhysRevLett.34.1530",
    journal = "Phys. Rev. Lett.",
    volume = "34",
    pages = "1530--1533",
    year = "1975"
}

@article{Kuznetsov:2023kss,
    author = "Kuznetsov, M. Yu. and Petrov, N. A. and Plokhikh, I. A. and Sotnikov, V. V.",
    title = "{Methods of machine learning for the analysis of cosmic rays mass composition with the KASCADE experiment~data}",
    eprint = "2311.06893",
    archivePrefix = "arXiv",
    primaryClass = "astro-ph.HE",
    doi = "10.1088/1748-0221/19/01/P01025",
    journal = "JINST",
    volume = "19",
    number = "01",
    pages = "P01025",
    year = "2024"
}

@article{ALEXANDREAS1993570,
title = {Point source search techniques in ultra high energy gamma ray astronomy},
journal = {Nuclear Instruments and Methods in Physics Research Section A: Accelerators, Spectrometers, Detectors and Associated Equipment},
volume = {328},
number = {3},
pages = {570-577},
year = {1993},
issn = {0168-9002},
doi = {https://doi.org/10.1016/0168-9002(93)90677-A},
url = {https://www.sciencedirect.com/science/article/pii/016890029390677A},
author = {D.E. Alexandreas and D. Berley and S. Biller and G.M. Dion and J.A. Goodman and T.J. Haines and C.M. Hoffman and E. Horch and X.-Q. Lu and C. Sinnis and G.B. Yodh and W. Zhang}
}

@article{Gorski_2005,
doi = {10.1086/427976},
url = {https://doi.org/10.1086/427976},
year = {2005},
month = {apr},
publisher = {},
volume = {622},
number = {2},
pages = {759},
author = {Górski, K. M. and Hivon, E. and Banday, A. J. and Wandelt, B. D. and Hansen, F. K. and Reinecke, M. and Bartelmann, M.},
title = {HEALPix: A Framework for High-Resolution Discretization and Fast Analysis of Data Distributed on the Sphere},
journal = {The Astrophysical Journal}
}

@article{Zonca2019, doi = {10.21105/joss.01298}, url = {https://doi.org/10.21105/joss.01298}, year = {2019}, publisher = {The Open Journal}, volume = {4}, number = {35}, pages = {1298}, author = {Zonca, Andrea and Singer, Leo P. and Lenz, Daniel and Reinecke, Martin and Rosset, Cyrille and Hivon, Eric and Gorski, Krzysztof M.}, title = {healpy: equal area pixelization and spherical harmonics transforms for data on the sphere in Python}, journal = {Journal of Open Source Software} }

@article{Gabici_2019,
   title={The origin of Galactic cosmic rays: Challenges to the standard paradigm},
   volume={28},
   ISSN={1793-6594},
   DOI={10.1142/s0218271819300222},
   number={15},
   journal={International Journal of Modern Physics D},
   publisher={World Scientific Pub Co Pte Ltd},
   author={Gabici, Stefano and Evoli, Carmelo and Gaggero, Daniele and Lipari, Paolo and Mertsch, Philipp and Orlando, Elena and Strong, Andrew and Vittino, Andrea},
   year={2019},
   month={Nov}, 
   pages={1930022} 
}

@article{LHAASO:2021gok,
    author = "Cao, Zhen and others",
    collaboration = "LHAASO",
    title = "{Ultrahigh-energy photons up to 1.4 petaelectronvolts from 12 $\gamma$-ray Galactic sources}",
    doi = "10.1038/s41586-021-03498-z",
    journal = "Nature",
    volume = "594",
    number = "7861",
    pages = "33--36",
    year = "2021"
}

@article{Antoni_2004,
doi = {10.1086/382039},
url = {https://doi.org/10.1086/382039},
year = {2004},
month = {apr},
publisher = {},
volume = {604},
number = {2},
pages = {687},
author = {Antoni, T. and Apel, W. D. and Badea, A. F. and Bekk, K. and Bercuci, A. and Blümer, H. and Bozdog, H. and Brancus, I. M. and Büttner, C. and Daumiller, K. and Doll, P. and Engel, R. and Engler, J. and Fessler, F. and Gils, H. J. and Glasstetter, R. and Haungs, A. and Heck, D. and Hörandel, J. R. and Kampert, K.-H. and Klages, H. O. and Maier, G. and Mathes, H. J. and Mayer, H. J. and Milke, J. and Müller, M. and Obenland, R. and Oehlschläger, J. and Ostapchenko, S. and Petcu, M. and Rebel, H. and Risse, A. and Risse, M. and Roth, M. and Schatz, G. and Schieler, H. and Scholz, J. and Thouw, T. and Ulrich, H. and van Buren, J. and Vardanyan, A. and Weindl, A. and Wochele, J. and Zabierowski, J. and (The KASCADE Collaboration)},
title = {Large-Scale Cosmic-Ray Anisotropy with KASCADE},
journal = {The Astrophysical Journal}
}

@article{Kostunin:2021eyp,
    author = "Kostunin, Dmitriy and Plokhikh, Ivan and Ahlers, Markus and Tokareva, Victoria and Lenok, Vladimir and Bezyazeekov, Pavel A. and Golovachev, Sergey and Sotnikov, Vladimir and Mullyadzhanov, Rustam and Sotnikova, E.",
    title = "{New insights from old cosmic rays: A novel analysis of archival KASCADE data}",
    eprint = "2108.03407",
    archivePrefix = "arXiv",
    primaryClass = "astro-ph.HE",
    doi = "10.22323/1.395.0319",
    journal = "PoS",
    volume = "ICRC2021",
    pages = "319",
    year = "2021"
}

@article{Ahlers:2016rox,
    author = "Ahlers, Markus and Mertsch, Philipp",
    title = "{Origin of Small-Scale Anisotropies in Galactic Cosmic Rays}",
    eprint = "1612.01873",
    archivePrefix = "arXiv",
    primaryClass = "astro-ph.HE",
    doi = "10.1016/j.ppnp.2017.01.004",
    journal = "Prog. Part. Nucl. Phys.",
    volume = "94",
    pages = "184--216",
    year = "2017"
}

@article{Lebedev_2026,
doi = {10.3847/1538-4357/ae510a},
url = {https://doi.org/10.3847/1538-4357/ae510a},
year = {2026},
month = {apr},
publisher = {The American Astronomical Society},
volume = {1001},
number = {2},
pages = {182},
author = {Lebedev, I. A. and Fedosimova, A. I. and Olimov, Kh. K. and Krassovitskiy, P. M. and Yerezhep, N. O. and Ibraimova, S. A. and Bondar, E. A.},
title = {Indications of Anisotropy of Cosmic-Ray Elemental Groups Based on KASCADE Data},
journal = {The Astrophysical Journal}
}

@article{He:2025N5,
  author = "He, Jiayin  and  Zhang, Yi  and  Yuan, Qiang",
  title = "{Observation of large-scale anisotropy of very high-energy cosmic-ray protons with LHAASO-KM2A}",
  doi = "10.22323/1.501.0286",
  journal = "PoS",
  year = 2025,
  volume = "ICRC2025",
  pages = "286"
}

\end{document}